\UseRawInputEncoding
\documentclass[]{spie}  %>>> use for US letter paper
\usepackage{amsmath,amsfonts,amssymb}
\usepackage{graphicx}
\usepackage[colorlinks=true, allcolors=blue]{hyperref}
\usepackage[table]{xcolor}
\usepackage{graphicx}
\usepackage{txfonts}
\usepackage{lscape}
\usepackage{hyperref}
\usepackage{comment}
\usepackage{xspace}

\usepackage{bm}
\usepackage{aas_macros}

\title{GrailQuest \& HERMES: Hunting for Gravitational Wave Electromagnetic Counterparts and Probing Space-Time Quantum Foam}

\author[a]{L.~Burderi}
\author[b]{T.~Di Salvo}
\author[a]{A.~Sanna}
\author[c]{F.~Fiore}
\author[a]{A.~Riggio}
\author[b]{A.~F.~Gambino}
\author[i]{F.~Amarilli}
\author[g]{L.~Amati}
\author[f]{F.~Ambrosino}
\author[j,k]{G.~Amelino-Camelia}
% ...

\affil[a]{Dipartimento di Fisica, Universit\`a degli Studi di Cagliari, SP Monserrato-Sestu km 0.7, I-09042 Monserrato, Italy}
\affil[b]{Dipartimento di Fisica e Chimica, Universit\`a degli Studi di Palermo, via Archirafi 36, I-90123 Palermo, Italy}
\affil[c]{INAF-OATS, Via G.B. Tiepolo 11, I-34143, Trieste, Italy}
% ...
\affil[f]{INAF-IAPS Rome, Via del Fosso del Cavaliere 100, I-00133, Italy}
\affil[g]{INAF-OAS Bologna, Via Gobetti 101, I-40129, Bologna, Italy}
% ...
\affil[i]{Fondazione Politecnico di Milano, Piazza Leonardo da Vinci, 32 20133 Milano, Italy}
\affil[j]{Dipartimento di Fisica Ettore Pancini, Universit\`a di Napoli ``Federico II'', and INFN, Sezione di Napoli, Complesso Univ. Monte S. Angelo, I-80126 Napoli, Italy}
\affil[k]{INFN, Sezione di Napoli, Complesso Univ. Monte S. Angelo, I-80126 Napoli, Italy}
% ...

\authorinfo{Further author information:\\Luciano Burderi: E-mail: burderi@dsf.unica.it, \\  Andrea Sanna: E-mail: andrea.sanna@dsf.unica.it}

\begin{document} 
\maketitle

\begin{abstract}
{\it GrailQuest} (Gamma-ray Astronomy International Laboratory for Quantum Exploration of Space-Time) is an ambitious astrophysical mission concept that uses a fleet of small satellites whose main objective is to search for a dispersion law for light propagation {\it in vacuo}.  
Within Quantum Gravity theories, different models for space-time quantization predict relative discrepancies of the speed of photons w.r.t. the speed of light that depend on the ratio of the photon energy to the Planck energy. This ratio is as small as $10^{-23}$ for photons in the $\gamma-{\rm ray}$ band ($100\,{\rm keV}$).
Therefore, to detect this effect, light must propagate over enormous distances and the experiment must have extraordinary sensitivity. 
Gamma-Ray Bursts, occurring at cosmological distances, could be used to detect this tiny signature of space-time granularity.
This can be obtained by coherently combine a huge number of small instruments distributed in space to act as a single detector of unprecedented effective area. 
This is the first example of high-energy distributed astronomy: a new concept of modular observatory of huge overall collecting area consisting in a fleet of small satellites in low orbits, with sub-microsecond time resolution and wide energy band (keV-MeV). 
The enormous number of collected photons will allow to effectively search these energy dependent delays. Moreover, {\it GrailQuest} will allow to perform temporal triangulation of impulsive events with arc-second positional accuracies: an extraordinary sensitive X-ray/Gamma all-sky monitor crucial for hunting the elusive electromagnetic counterparts of Gravitational Waves, that will play a paramount role in the future of \emph{Multi-messenger Astronomy}. 
A pathfinder of {\it GrailQuest} is already under development through the {\it HERMES} (High Energy Rapid Modular Ensemble of Satellites) project: a fleet of six 3U cube-sats to be launched by the end of 2022. 
\end{abstract}

\keywords{Gamma-Ray Bursts, X-rays, CubeSats, nano-satellites, temporal triangulation, Quantum gravity, Gravitational Wave counterparts, all-sky monitor, Temporal triangulation}

\section{INTRODUCTION: Two compelling (astro)$-$physical problems for the next decades}

In this paper we review a new concept of modular observatory for high-energy astronomy from space. 
This new type of space observatory is based on the principle of Distributed Astronomy in which the overall capabilities of the instrument arise from the simultaneous use of very small and simple sub-units. 
This principle allows the construction, in space, of instruments with an enormous overall effective area. 
Distributed Astronomy will allow to tackle two of the most compelling (astro)-physical problems of the next decades:
\begin{itemize}
\item[]
{\it i)} {\it The development of multi-messenger astronomy from infancy to maturity}.
The construction of a very high sensitivity all-sky monitor for the accurate localization of transient events in the X-/gamma-ray band is mandatory for the fast identification and localization of the electromagnetic counterparts of some Gravitational Wave Events (GWE).
\item[] 
{\it ii)} {\it To probe Space-Time granularity down to the Planck scale ($\ell_{\rm PLANCK} = 1.6 \times 10^{-33}\; {\rm cm}$)}. 
The realization of a huge effective area X-/gamma-ray telescope allows to perform an ambitious Quantum Gravity experiment: to search for a dispersion law for light in vacuo, that linearly depends on the ratio between photon energy and Planck energy. 
\end{itemize}

\subsection{The birth of Multi$-$Messenger Astronomy and the Multi$-$Messenger Astronomy Paradox}
\begin{figure}[h!]
\begin{center}
\begin{tabular}{c}
\includegraphics[height=13cm]{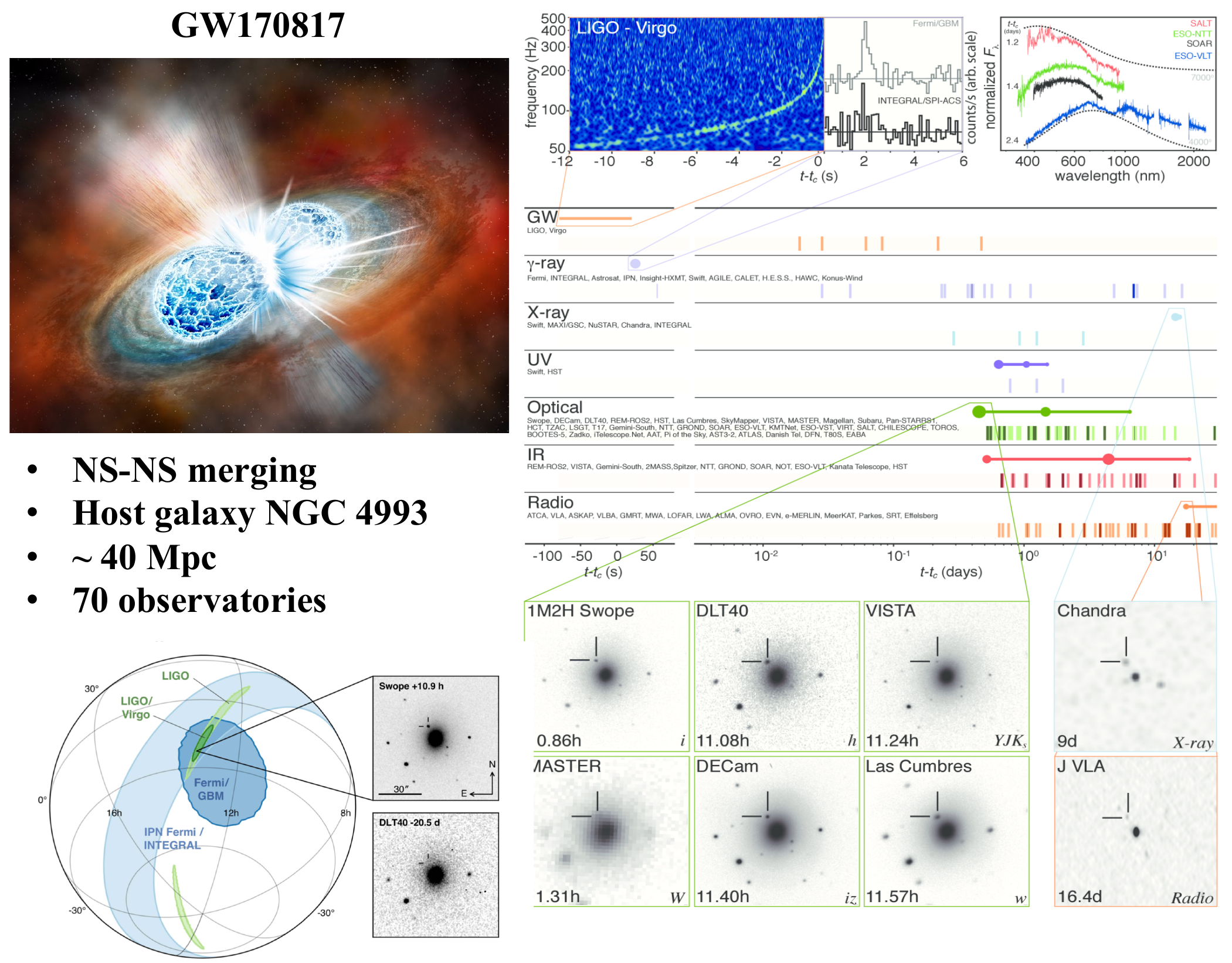} 
\end{tabular}
\end{center}
\caption[example]
%>>>> use \label inside caption to get Fig. number with \ref{}
{ \label{fig:multimessenger} 
The birth of multi-messenger astronomy: timeline of multi-wavelength detection of GW170817/GRB170817A. Figures from [\citenum{Abbot_mu_2017}].
}
\end{figure}

In August 2017, the first neutron star-neutron star (NS-NS) merger has been discovered by LIGO/Virgo gravitational wave interferometers [\citenum{GW170817}]. 
A short Gamma Ray Burst (GRB 170817A), seen off jet-axis, was detected 1.7 seconds after the event, first by the Gamma-ray Burst Monitor (GBM) on board of the Fermi satellite, and later by INTEGRAL and other observatories [\citenum{GW170817_GW,Troja17}]. 
The intersection of the sky error box of LIGO/Virgo with that of GBM led to the first identification of an optical transient associated to a short GRB and a GWE, opening, {\it de facto}, the era of multi-messenger astronomy [\citenum{Abbot_mu_2017}]. The timeline of the multi-wavelength detection is shown in Figure~\ref{fig:multimessenger}.

\begin{figure}[h!]
\begin{center}
\begin{tabular}{c}
\includegraphics[height=9cm]{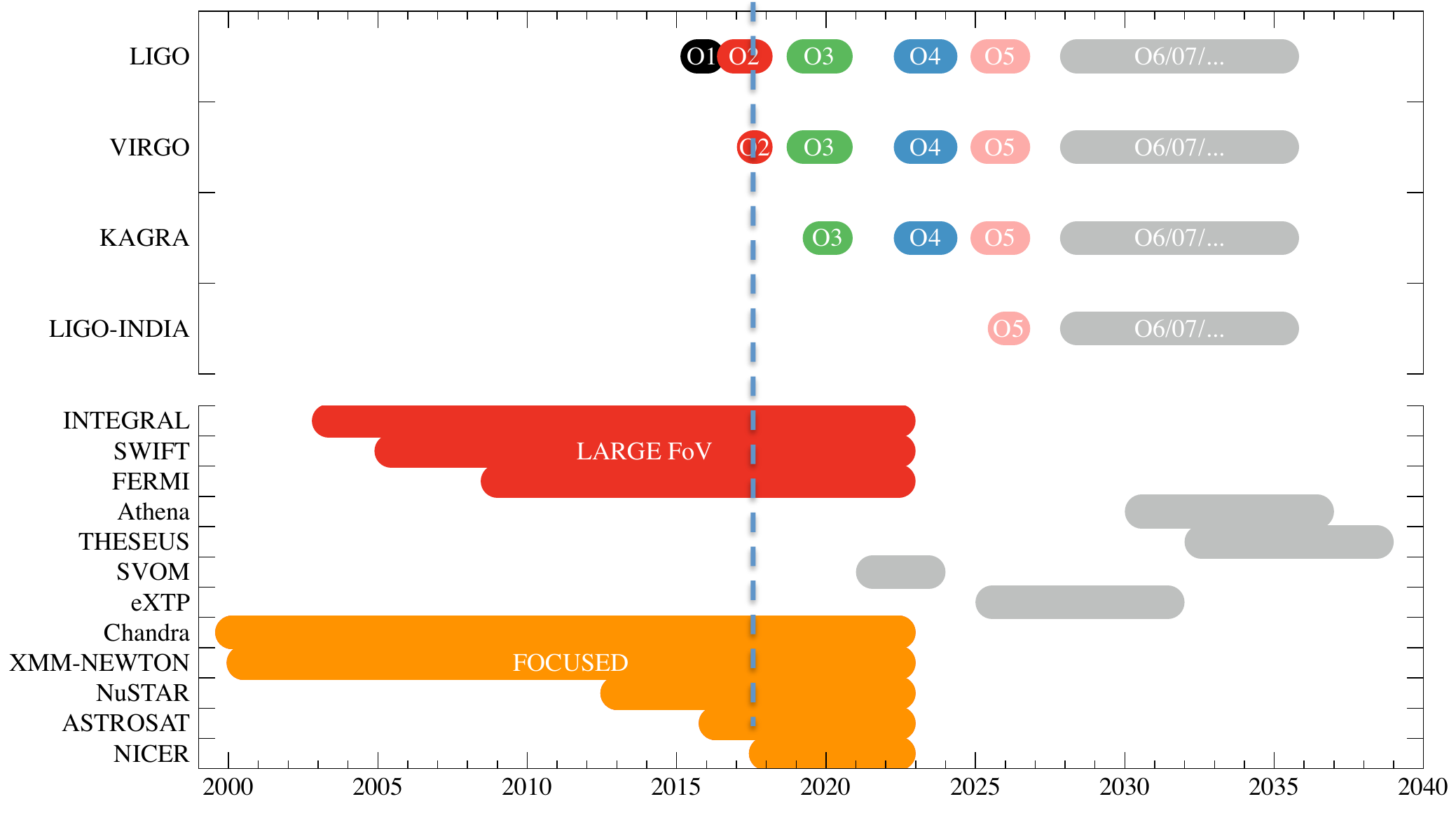} 
\end{tabular}
\end{center}
\caption[example] 
%>>>> use \label inside caption to get Fig. number with \ref{}
{ \label{fig:gwandxobs} 
Schematic view of the operative schedule of Gravitational Wave Interferometers and X-/gamma-ray Observatories in the period 2000-2040.}
\end{figure}

\noindent Gravitational-wave interferometers are expected to both expand in number and increase in sensitivity over the next 5-10 years. The first two observing runs included both of the LIGO interferometers as well as the French-Italian Virgo interferometer. The third observing run will likely see the inclusion of a fourth site in Japan, KAGRA [\citenum{Castelvecchi19}]. A fifth site in India, LIGO-India is in the planning, with a 2024 commissioning date [\citenum{Priyadarshini16}]. By the middle of the decade, a five-site world-wide network will be operational, with a detection horizon of approximately 300 Mega-parsecs for binary NS mergers. On the other hand, most of the present large Field of View X-/gamma-ray observatories are expected to end operations by 2025. The situation is summarized in Figure~\ref{fig:gwandxobs}.
\begin{table*}[h!]
\begin{center}
Expected rate of detectable Gravitational Wave Events 
\vskip 0.2cm
\bf{
\begin{tabular}{ll}
\hline
\hline
\\
${\rm Objects}$ & ${\rm Rate}\, ({\rm Gpc^{-3}yr^{-1}})$ \\
% & ${\rm Gpc^{-3}yr^{-1}}$ \\
 & \\
${\rm Short-GRBs/kilonovae}$ & $1540$ \\
\\
${\rm Long-GRBs/Hypernovae}$ & $225$ \\
\\
${\rm SuperLuminous SNe}$ & $100$ \\
\\
${\rm CC-SNe}$ & ${\rm about\, one\, event}$ \\
& ${\rm per\, year\, within}$ \\
& $25\, {\rm Mpc}$ \\
\hline
\hline
\end{tabular}
}
\end{center}
\caption{\label{table:0} 
}
\end{table*}
\noindent While GW170817 was detected with relative ease due to its close by distance (40 Mega--pc), future detections may not be so easy. This is particularly true as the sensitivity horizon of Gravitational Wave detectors spreads out to hundreds of Mega--parsecs, allowing the detection of few NS--NS merger events per year. 
%INIZIO PARTE PROPOSTA DA MASSIMO DELLA VALLE %%%%%%%%%%%%%%%%%%%%%%%%%%%%%%%%%%%%%%%%%
Indeed for Gravitational Wave detectors we have the following estimates: 1540 (+3200 -1220) Gpc-3 yr-1 [\citenum{GW170817}] for NS-NS merging and therefore Short-duration GRBs + Kilonovae rate.
Long-duration GRBs+HNe: 225 events Gpc-3/yr  [\citenum{Guetta2007}]. If we search for synergy with electromagnetic observations we should rescale this number for a beaming  angle of 4- 8 deg. So the electromagnetic rate  of long GRBs drastically decreases to about 1 GRB per Gpc-3/yr. 
Other catastrophic events like Super Luminous SNe producing rapidly rotating black holes. The expected rate is circa 100 events per Gp3 per year.
Finally we should include among the potential targets very nearby Core-Collapse Supernovae. The estimated frequency of occurrence for these objects is about 70,000 Gpc-3 yr-1 [\citenum{Cappellaro2015}].
Obviously since far less energy is emitted in GWs by CC-SNe than GRBs (the former phenomena involve -as final product- the formation of neutron stars which are well below the energy reservoir in angular momentum of black holes of similar mass, given their limited rotational energy), they can be detected as GW sources only within the very local universe, likely within the Virgo circle of ~ 20 Mpc, which implies about 1 event /yr.
These estimates are summarized in the Table \ref{table:0}. 
%FINE PARTE PROPOSTA DA MASSIMO DELLA VALLE %%%%%%%%%%%%%%%%%%%%%%%%%%%%%%%%%%%%%%%%%

Indeed, Fermi-GBM would not have detected the counterpart of an event like GW170817 at distances greater than 60 Mega-parsecs.
We therefore need an all-sky monitor with an area at least 10 to 100 times larger than GBM for letting multi-messenger astronomy to develop from infancy (one event, GW170817/GRB170817A, detected up to date) to full maturity.

\subsection{Gamma-Ray Burst in a nutshell: phenomenology and theoretical models}

\begin{figure}[h!]
\begin{center}
\begin{tabular}{cc}
\includegraphics[height=7cm]{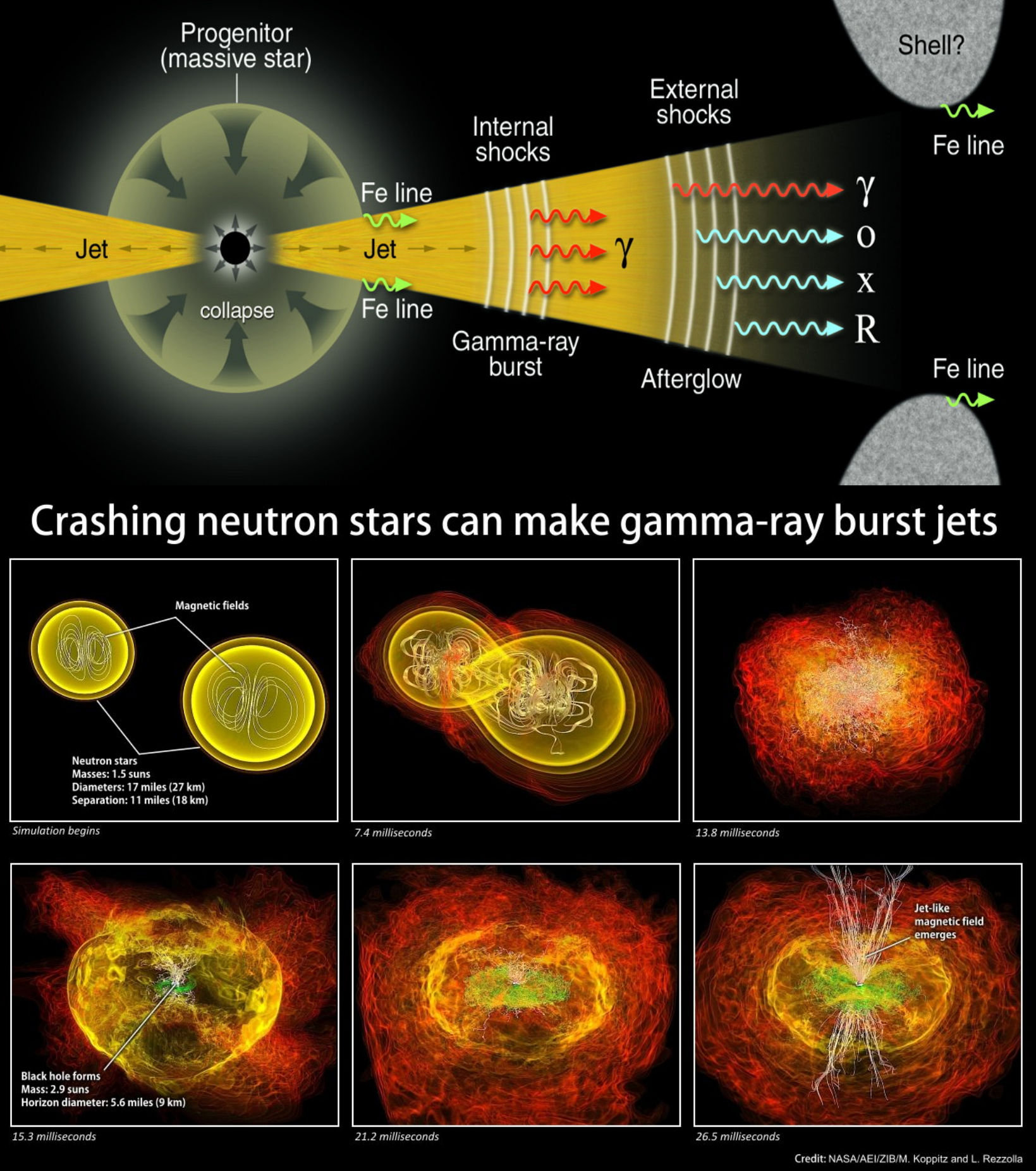} &
\includegraphics[height=7cm]{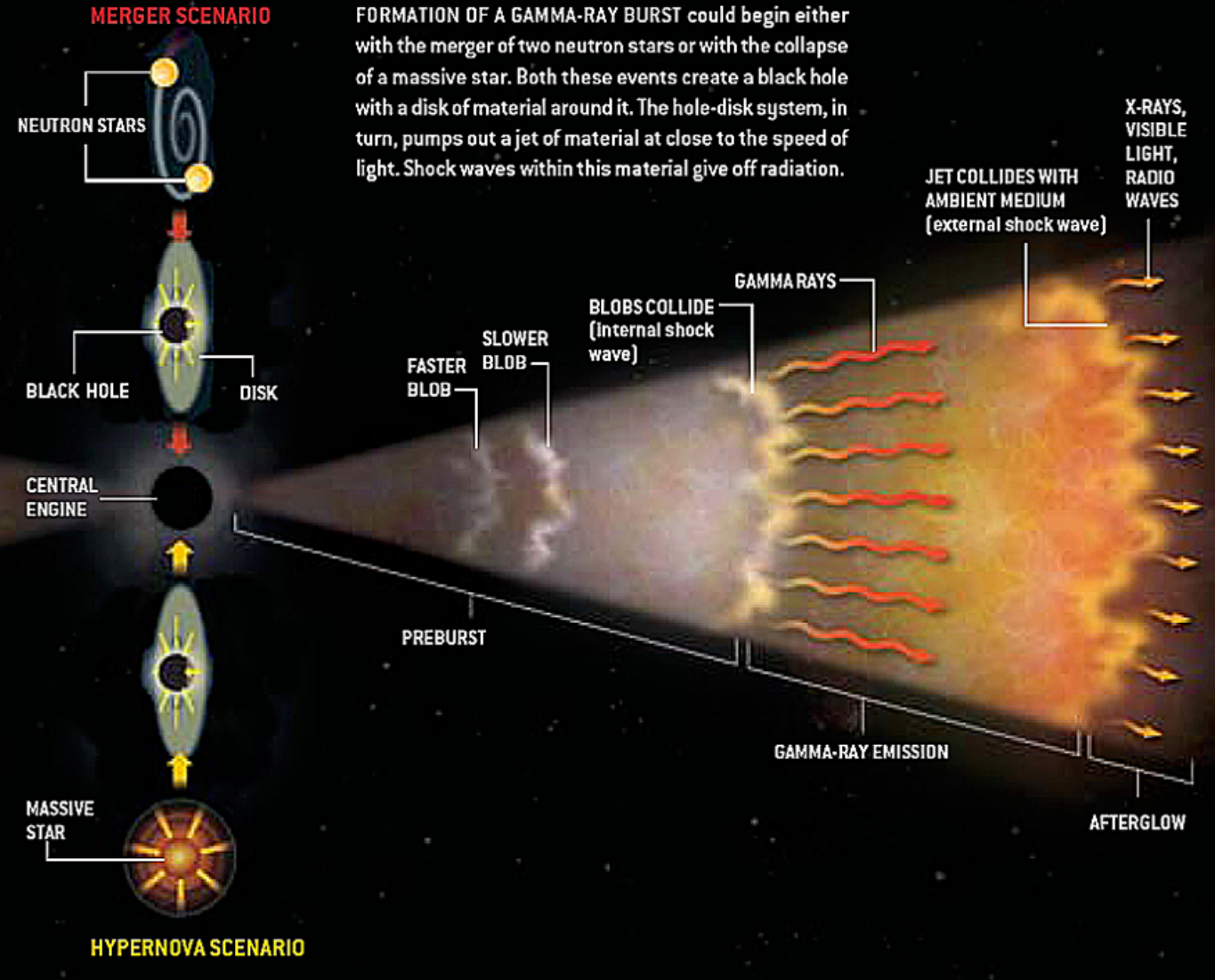} \\
Panel a) & Panel b) \\
\end{tabular}
\end{center}
\caption[example] 
%>>>> use \label inside caption to get Fig. number with \ref{}
{\label{fig:fireball} \emph{Panel a)} Schematic representation of a Hypernova (top) and of the final phases of a NS-NS merger. 
\emph{Panel b)} The {\it Fireball model} for GRBs. Credits to NASA / ALBERT EINSTEIN INSTITUTE / ZUSE INSTITUTE BERLIN / M. KOPPITZ AND L. REZZOLLA and JUAN VELASCO)
}
\end{figure}

In the following we summarize the GRBs phenomenology, highlighting the features most relevant for the present paper:
\begin{itemize}
\item[]
{\it i)} sudden and unpredictable bursts of hard-X / soft gamma rays with huge flux
\item[] 
{\it ii)}  most of the flux detected from 10-20 keV up to 10 MeV
\item[] 
{\it iii)} occurrence rate of very bright GRBs ($25\, {\rm counts\, cm^{-2} s^{-2}}$ in the $20-300\, {\rm keV}$ band) is $\sim 3{\rm yr^{-1}}$. The outlier monster GRB130427A reached a record flux of $\sim 160\, {\rm counts\, cm^{-2} s^{-2}}$
\item[] 
{\it iv)} presence of a bimodal distribution of duration $0.1-1.0$ s (Short GRB) and  $10-100$ s (Long GRBs) (see e.g. [\citenum{Kouveliotou1993,Zhang2008}])
\item[] 
{\it v)} measured rate (by an all-sky experiments on a LEO satellites) $\sim 0.8/$day [\citenum{vonKienlin20}] (estimated true rate $\sim 2\div3$/ day) 
\item[] 
{\it vi)} Long and Short GRB with millisecond time variability are about $40\%$ of bright GRBs. There is evidence of sub-millisecond variability in some GRBs (see e.g. [\citenum{Walker00,Bhat12,MacLachlan13}])
\item[] 
{\it vii)} presence of an afterglow in X-rays,  UV, optical, IR, millimeter, radio (see e.g. [\citenum{Galama2000}])
\item[] 
{\it viii)} redshift measured in afterglow and host galaxies (see e.g. [\citenum{Schilling02}] for a review)
\item[] 
{\it ix)} cosmological origin:spatial isotropy and distance measured from redshifts in afterglows and host galaxies (see e.g. [\citenum{Schilling02}] for a review)
\end{itemize} 
Proposed GRB progenitors are the collapse of a massive star (Hypernova model for Long GRBs) [\citenum{Woosle1993,MacFadyen2001}] and the merger (because of gravitational wave emission) of two NSs (Kilonova model for Short GRBs) [\citenum{Gehrels2005,Barthelmy2005,Fox2005}]. Both these events create a black hole with a transient disk of material around it that pumps out a jet of material at a speed close to the speed of light. 
In the so called {\it Fireball model}, a compact source releases a few $10^{51}$ ergs within tens of seconds in a 10 km radius region. Regardless of the form of energy initially released, a quasi-thermal equilibrium between radiation and matter is reached. This electron-positron plasma, clumped in thin shells and opaque to radiation, accelerates to relativistic velocities with Lorentz factors of 
$\gamma = [ 1 - (v/c)^2 ]^{-1/2} = 100 \div 1000$ (where $v$ is the speed of the shell and $c$ is the speed of light) until a considerable fraction of the initial energy has been converted into bulk kinetic energy. The plasma is collimated into a jet of few tens degrees opening angle. Multiple collision of relativistic shells of slightly different Lorentz factors cause the prompt emission through synchrotron radiation and inverse Compton scattering. Furthermore, shock of outer shells with interstellar medium originates the so-called afterglow that generates radiations from X-rays down to radio. Long and Short GRBs progenitors and details of the {\it Fireball model} are shown in Figure~\ref{fig:fireball}

\section{Distributed Astronomy in a nutshell: HERMES and GrailQuest missions in this context}

Distributed Astronomy is an effective way to build an all-sky monitor of excellent sensitivity to locate in the sky with great accuracy and study fast variability of high-energy transient, and for continuous monitoring of periodic sources. Each detector has a half-sky field of view, and localization capabilities are obtained by temporal triangulation of an impulsive or periodic signal detected by a network of detectors distributed in space (see section \ref{sec:trig}, below). Nano-satellites can host 100 cm$^2$ detector in the keV-MeV range. The advantages of using fleets of nano-satellites reside in the modularity of the experiment, allowing for:
\begin{itemize}
\item[]
{\it i)} build-up a huge overall effective area. 
\item[]
{\it ii)} mass-production and subsequent cost reduction.
\item[]
{\it iii)} quick development and continuous upgrade of the detectors.
\end{itemize}
{\it HERMES} (High Energy Rapid Modular Ensemble of Satellite) is a pathfinder experiment consisting of a fleet of six 3U nano-satellites in Low Earth Orbit to be launched by the end of 2022. \\
On the other hand, {\it GrailQuest} (Gamma Ray Astronomy International Laboratory for Quantum Exploration of Space-Time) is a mission concept including a vast fleet of hundreds/thousands of satellites proposed for the Voyage 2050 - long term plan in the ESA science program. 

\subsection{Principles of temporal triangulation}
\label{sec:trig}

\begin{figure}[h!]
\begin{center}
\begin{tabular}{c}
\includegraphics[height=6cm]{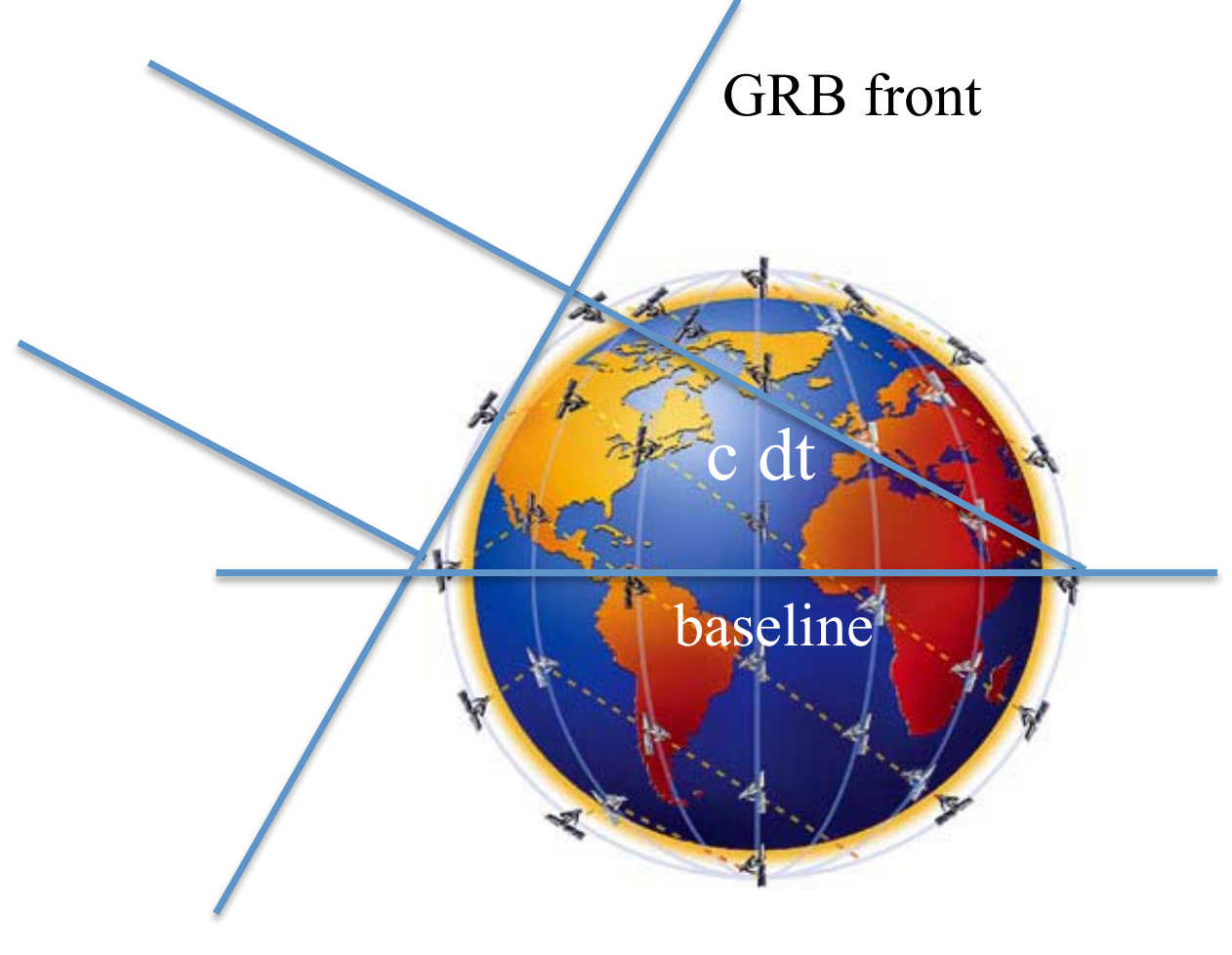} 
\end{tabular}
\end{center}
\caption[example] 
%>>>> use \label inside caption to get Fig. number with \ref{}
{ \label{fig:triang} 
Temporal triangulation in a bi-dimensional case.}
\end{figure}

A very promising technique for accurate localization of transient astrophysical sources is the so-called {\it Temporal Triangulation}. The idea is simple and robust, as outlined in the following. Let us represent the transient event as a narrow -in time- wavefront (pulse) traveling in a given direction. Let us displace a network of detectors in space. The narrow wavefront will hit the detectors of the network at different times that depend on the spatial position of each detector and the direction of the wavefront. Consider the simplest case of three detectors ({\it e.g.} A, B, C) displaced on a plane on the vertex of an equilateral triangle, $D$ being the diameter of the circumscribed circumference. The three Time of Arrivals (ToAs, hereafter) of the wavefront on each detector define the absolute ToA of the signal ({\it e.g.} the ToA on the detector A, chosen to represent reference of the time axis) and two delays ({\it i.e.} the ToAs on B and C w.r.t. the ToA on A) that uniquely define the direction of the source in the sky ({\it e.g.} through a system of two equations in the two unknowns $\alpha$ and $\delta$, representing its celestial coordinates). 
A simplified bi-dimensional model is shown in Figure~\ref{fig:triang}
In a broad sense, this method constrains the position of the source in the sky in an analogous way in which the diffraction limit of optical devices constrains the angular resolution $\Delta \theta$. Indeed, in the diffraction limit, the accuracy is limited by the capability of the optical device to be sensitive to phase differences that are of the order of $\phi = \pm \lambda/2$, since a difference in phase modulus of $\lambda/2$ implies the variation from constructive to destructive interference (hereafter we measure phases in units of wavelength). Given a baseline $D$ of the order of the size of the collector of the waves ({\it e.g.} the diameter of the mirror, for optical telescopes) simple trigonometric considerations imply $\Delta \theta \sim \lambda/D$. In perfect analogy, in performing a temporal triangulation, we use the ToA of the transient signal (pulse) as a proxy of the phases of the electromagnetic wave. Given an uncertainty $\sigma_{\Delta t}$ in the delays of the ToA of two detectors, the associated uncertainty in the spatial distance travelled by the pulse, $\Delta s = c \sigma_{\Delta t}$, correspond to the limit in sensitivity to phase differences of optical devices. Therefore the "diffraction limit" of temporal triangulation techniques is given by $\Delta \theta \sim c \sigma_{\Delta t}/D$, where $D$ is typical distance between the detectors. 

Pursuing this analogy further, we can understand the paramount difference between different interferometric techniques. Direct interferometry is the technique adopted in some optical and radio arrays of telescopes, {\it e.g.} the optical telescopes of Very Large Telescope, VLT, in the Atacama Desert, Chile, (see {\it e.g.} the recent paper on the first direct detection of an exoplanet by optical interferometry [\citenum{Gravitycolletal19}]) or the radio telescopes of the Very Large Array, VLA, in Socorro, New Mexico (see {\it e.g.} \url{https://public.nrao.edu/telescopes/vla/vla-basics}). In these cases, the optical and radio waves are allowed to interfere directly through waveguides that convey them appropriately. On the other hand, off-line radio interferometry is adopted for Earth-sized array of radio telescopes, {\it e.g.} those of Very Large Baseline Interferometer, VLBI, network (see {\it e.g.} \url{https://www.nrao.edu/index.php/about/facilities/vlba}) or in processing data from the different radio and millimeter telescopes of the Event Horizon Telescope (EHT) project, an Earth size telescope array consisting of a global network of radio telescopes with a angular resolution sufficient to resolve the event horizon of a supermassive black hole. In 2019 the EHT Collaboration published the first image of the region surrounding the event horizon of the supermassive black hole at the center of galaxy Messier 87 [\citenum{Ehtcoll19}]. In these cases, the ToA of the phases of the waves were recorded at each detector and, subsequently, the phases were combined with numerical codes. 
Direct optical or radio interferometry are complicated versions of optical devices in which scientists content themselves with knowing the outcome of the interference phenomenon and not the separate values of the phases of the waves in the individual detectors. We want to observe that for an electromagnetic signal that is revealed through the detection of each single quanta on different detectors of different devices, the interference pattern, and therefore the phase of the wave, is lost in the detection process. This is certainly the case for optical light and therefore direct interferometry is, at moment, the only viable technique in this case.
On the other hand, if the electromagnetic signal can be treated classically, it is possible to record the variable (w.r.t. time) amplitude of, {\it e.g.}, the electric vector of the wave and therefore the phases can be reconstructed a posteriori. This is the case of VLBI or EHT observations where off-line radio interferometry is applicable.

Temporal triangulation is the analog of off-line radio interferometry where temporal delays play the role of phase differences. It is straightforward that increasing the number of detectors from three to $N_{\rm DET} >3$, the number of independent delays adopted to determine the two quantities that define the position of a source in the sky ({\it e.g.} $\alpha$ and $\delta$, as before) is overdetermined and equal to $N_{\rm IND} = (N_{\rm DET} -1)-2$, and the accuracy can be treated in a statistical way. For $N_{\rm DET} > 3$, these statistical errors scales as:
\begin{equation} 
\label{eq:posacc} 
\sigma_{\alpha \, {\rm STAT}} \sim  \sigma_{\delta \, {\rm STAT}} \sim c \sigma_{\Delta t}D^{-1}(N_{\rm DET} - 3)^{-1/2}.
\end{equation}

To fully appreciate the potential of temporal triangulation, it is instructive to compare the resolving power of the VLA, 27 radio telescopes capable of moving on the radii of a circle with a maximum diameter of 40 km, which operates in the radio band at wavelengths between 0.7 and 400 cm, with that of a configuration of detectors for high-energy photons, in the keV-MeV band, arranged at distances comparable to those of the Earth-Moon system Lagrangian points, which we will call Lagrange System, in this example. For the VLA, we adopt an average diameter $D_{\rm VLA} = 20\,{\rm km}$, an average wavelength $\lambda = 20\,{\rm km}$, while for the Lagrange System we adopt an average diameter of $D_{\rm LAG} = 8 \times 10^5 \, {\rm km}$ and $\sigma_{\Delta t} = 0.2 {\rm ms}$, for the uncertainty in the delays of the ToA of two detectors (see below). The diffraction limit of the VLA is $\Delta \theta \sim \lambda/D = 20\,{\rm cm}/20\,{\rm km} \sim 2\,{\rm arcsec}$. The "diffraction limit" of the Lagrange System is $\Delta \theta \sim c \sigma_{\Delta t}D^{-1} = 6 \times 10^6\,{\rm cm}/8 \times 10^{10}\,{\rm cm} \sim 15\,{\rm arcsec}$. This shows that a temporal triangulation system displaced in space around the Earth and the Moon is capable to locate high-energy transients with a positional accuracy only slightly worse than the VLA.

%\subsection{Monte-Carlo simulations of a true long GRB and cross-correlation analysis of light-curves from two detectors}
\subsubsection{GRB Simulations and cross-correlation analysis}
\label{sec:montecarlo}
\begin{figure}[h!]
\begin{center}
\begin{tabular}{cc}
\includegraphics[height=6cm]{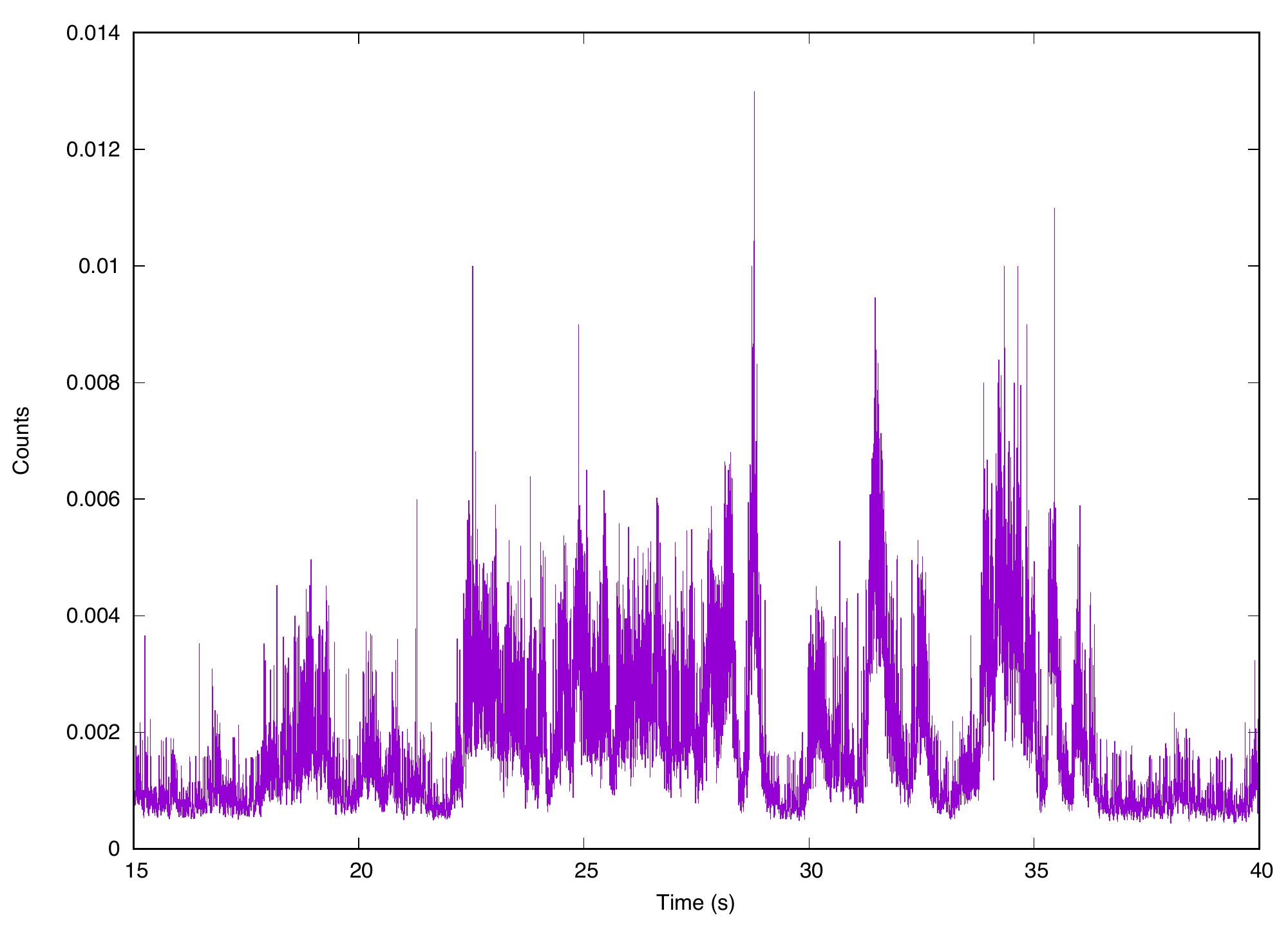} &
\includegraphics[height=6cm]{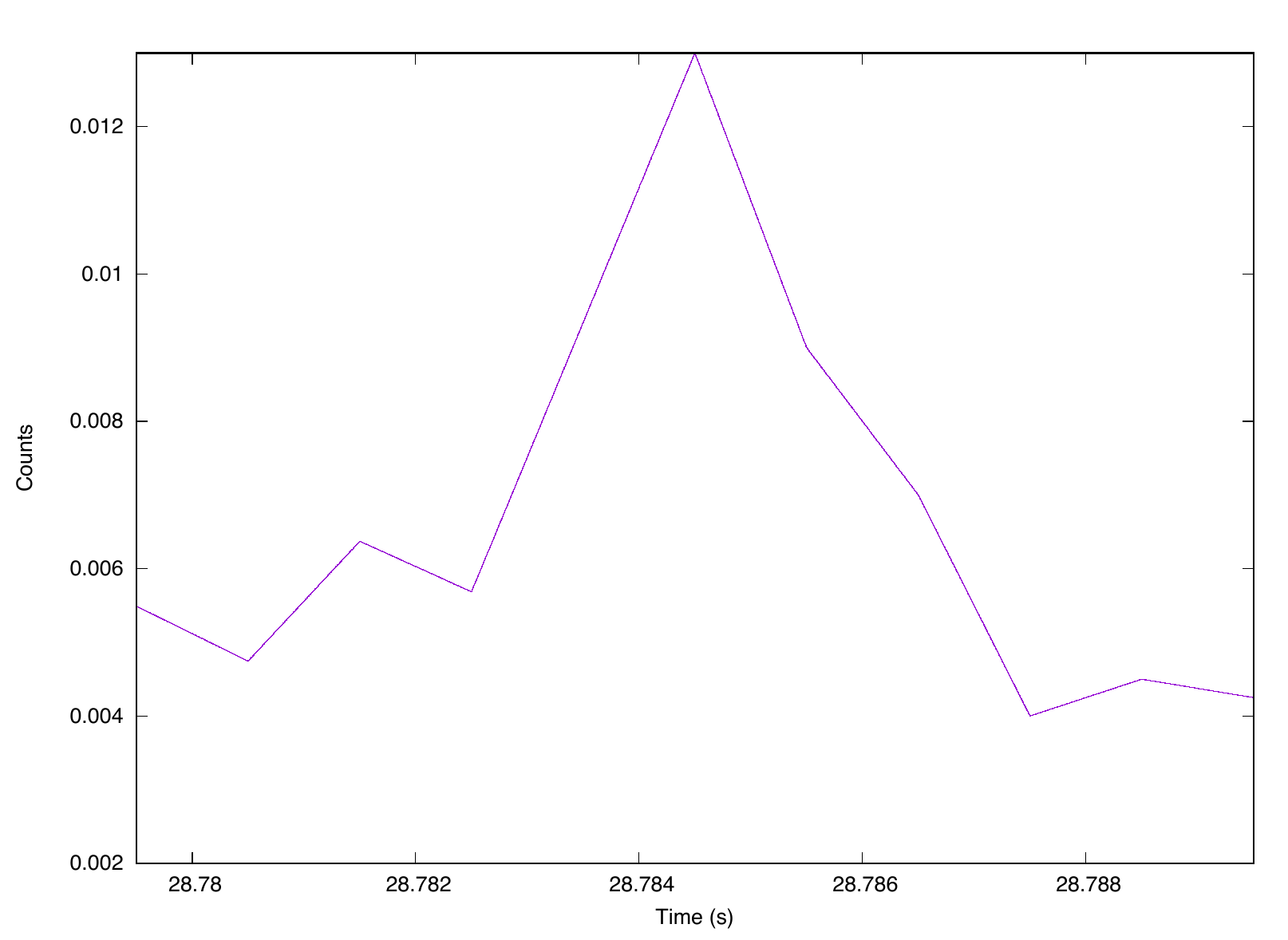} \\
Panel a) & Panel b) \\
\end{tabular}
\end{center}
\caption[example] 
%>>>> use \label inside caption to get Fig. number with \ref{}
{\label{fig:grb} GRB130502327 observed by Fermi-GBM.
\emph{Panel a)}  GRB light-curve.
\emph{Panel b)} Template at one millisecond resolution (detail of the main peak).
}
\end{figure}

Here we describe the approach adopted to exploit temporal triangulation capabilities to investigate GRBs.
The light-curve of a bright long GRB observed by Fermi-GBM is shown in Figure~\ref{fig:grb}, left panel.
The bright Long GRB lasted for $\Delta t_{\rm GRB} = 40 \, {\rm s}$, with an average flux in the 50-300 keV energy band of 
$\phi_{\rm GRB} = 6.5 \, {\rm photons}/{\rm s}/{\rm cm}^2$,
and a background flux of $\phi_{\rm BCK} = 2.8 \, {\rm photons}/{\rm s}/{\rm cm}^2$. Moreover, the GRB is characterized by variability on timescale of the order of $\sim 5 \, {\rm ms}$.

\begin{figure}[h!]
\begin{center}
\begin{tabular}{cc}
\includegraphics[height=5.5cm]{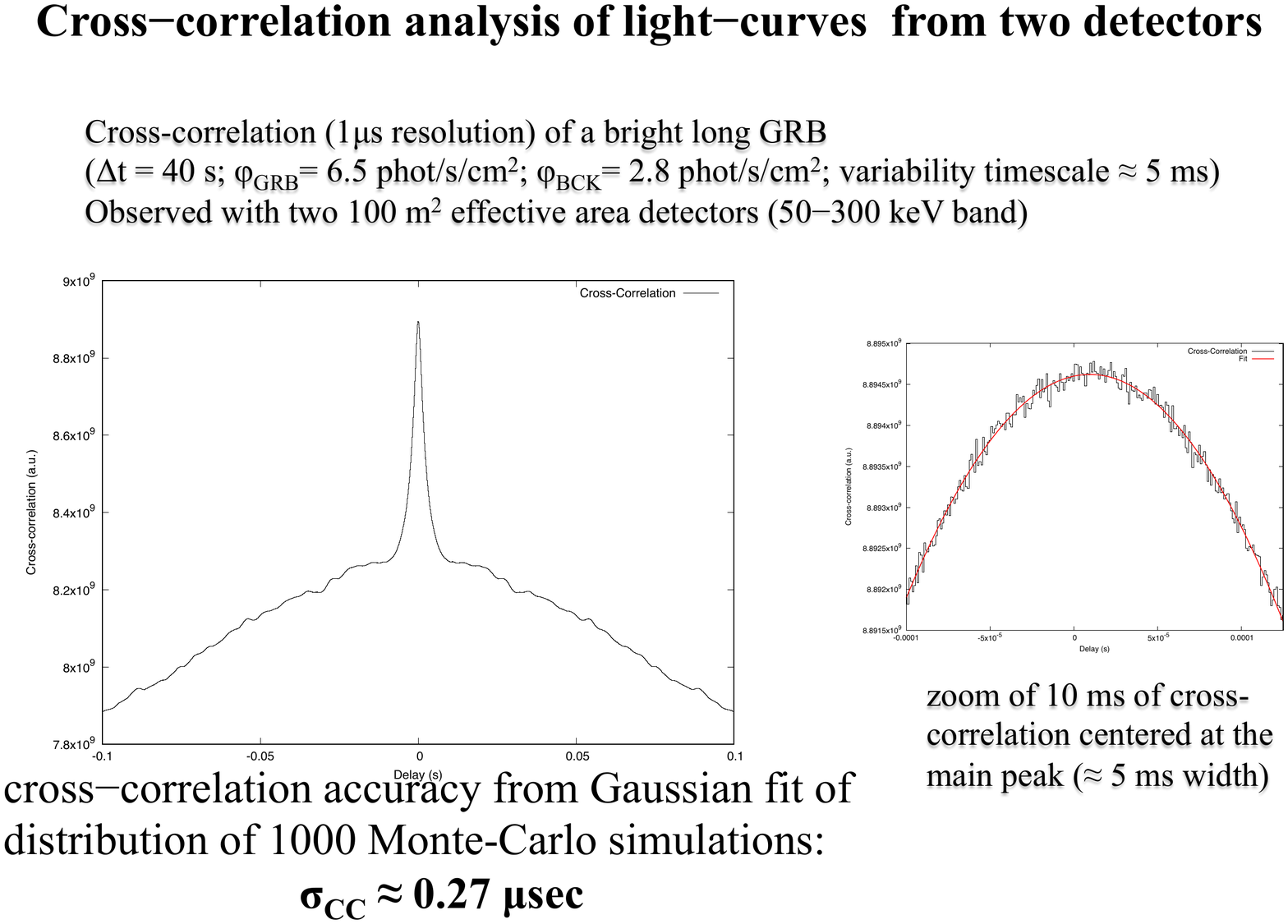} &
\includegraphics[height=5.5cm]{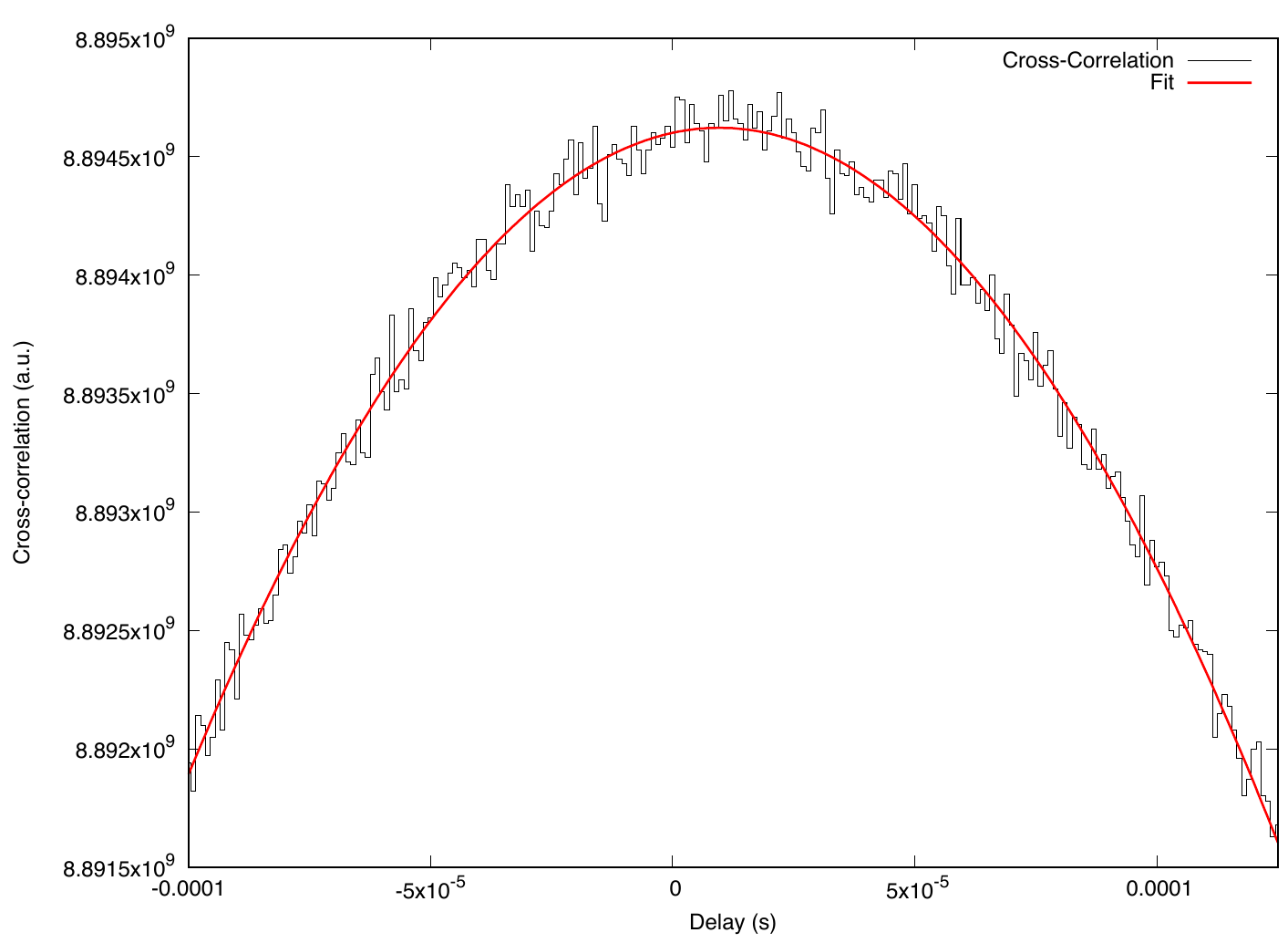} \\
Panel a) & Panel b) \\
\end{tabular}
\end{center}
\caption[example] 
%>>>> use \label inside caption to get Fig. number with \ref{}
{\label{fig:cross} Cross--correlation analysis of a simulated GRB seen by two identical detectors.
\emph{Panel a)} Cross--correlation function.
\emph{Panel b)} Detail of the cross--correlation function at $1 \mu{\rm s}$ around the main peak.
}
\end{figure}

\begin{figure}[h!]
\begin{center}
\begin{tabular}{c}
\includegraphics[height=14cm]{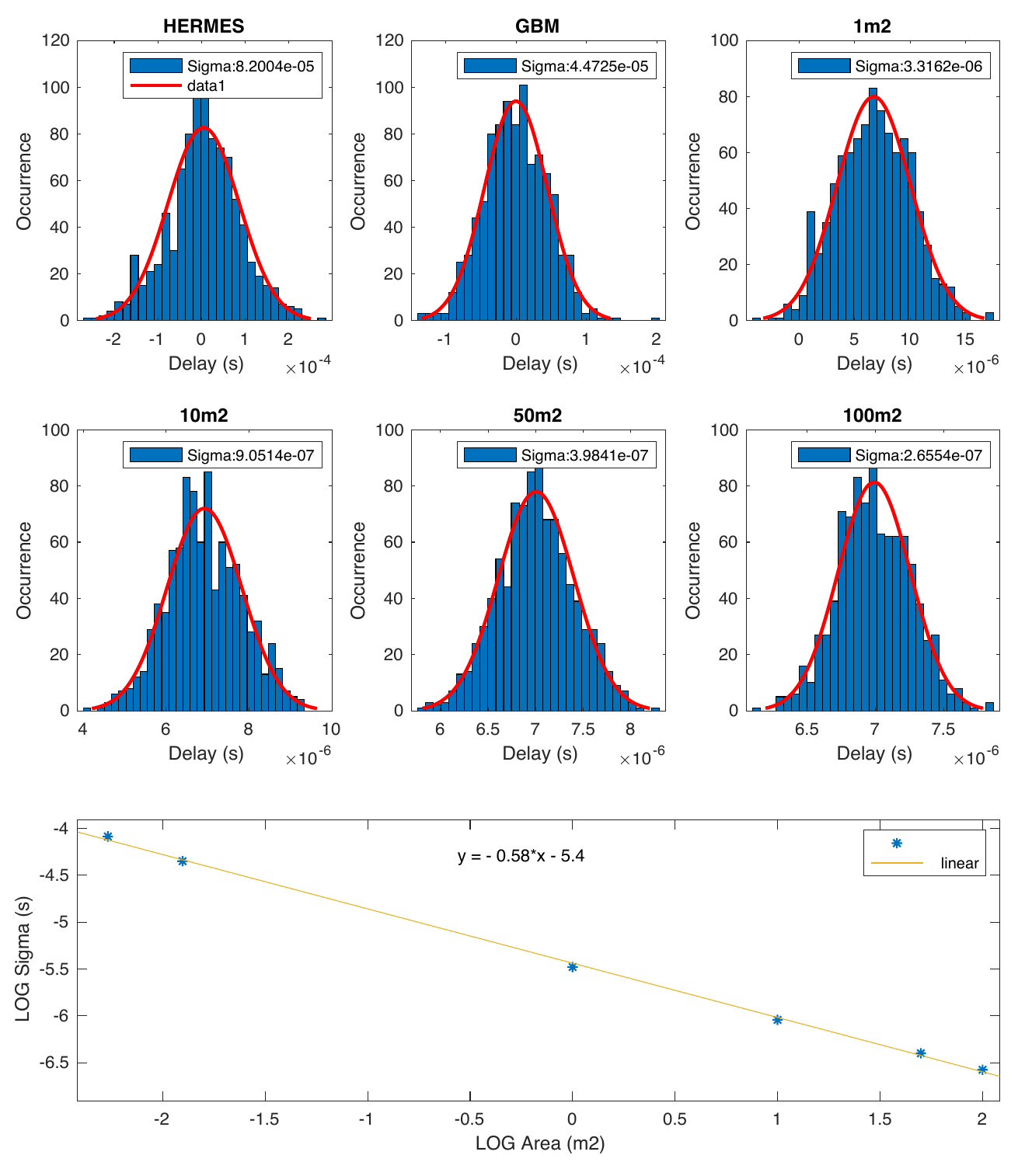} 
\end{tabular}
\end{center}
\caption[example] 
%>>>> use \label inside caption to get Fig. number with \ref{}
{ \label{fig:distr} 
Accuracy in cross correlation derived from Monte-Carlo simulations of GRBs.
\emph{Top panel)} Distributions of the delays obtained from cross-correlation analysis between 1000 pairs of simulated light-curves of identical detectors, for different effective areas, $56\, {\rm cm}^2$ (HERMES), $125\, {\rm cm}^2$ (Fermi-GBM), $1\, {\rm m}^2$, $10\, {\rm m}^2$, $50\, {\rm m}^2$, and $100\, {\rm m}^2$. 
\emph{Bottom panel)} Logarithmic plot of the one sigma delay accuracy as a function of the effective area and best linear fit.}
\end{figure}

Starting from this, we derived a template with millisecond resolution (see [\citenum{Sanna2020}] for more details on the method used to create the template). Figure~\ref{fig:grb}, right panel, shows the detail of the main peak of the GRB template where the timescale of the fast variability is about 5 ms. 
Using Monte-Carlo simulations, we generated light-curves as seen by detectors of different effective areas located in different positions of space. We performed cross-correlation analysis between pairs of simulated GRB with the aim to investigate the capability to reconstruct time delays between the observed signals. As an example, the cross-correlation function at $1 \mu{\rm s}$ resolution for a pair of detectors of $100\, {\rm m}^2$ area is shown in Figure~\ref{fig:cross}, left panel. The right panel shows the detail of the cross-correlation function around the peak and the best fit Gaussian. To determine a reliable estimation of the accuracy achievable using cross-correlation analysis, we repeated the procedure described 1000 times, and we then fitted the distribution with a Gaussian model, from which we estimated an accuracy of $0.27\, \mu{\rm s}$.
 Distributions for different effective areas, $56\, {\rm cm}^2$ (HERMES), $125\, {\rm cm}^2$ (Fermi-GBM), $1\, {\rm m}^2$, $10\, {\rm m}^2$, $50\, {\rm m}^2$, and $100\, {\rm m}^2$, are shown in the six panels of Figure~\ref{fig:distr}, top panel. 
The bottom panel shows the one sigma delay accuracy as a function of the effective area. The accuracy scales as the inverse of the effective area $A$ to the power of 0.6, close but slightly better than the theoretical lower limit of 0.5 (grossly derived from counting statistics). 

\begin{figure}[h!]
\begin{center}
\begin{tabular}{c}
\includegraphics[height=8cm]{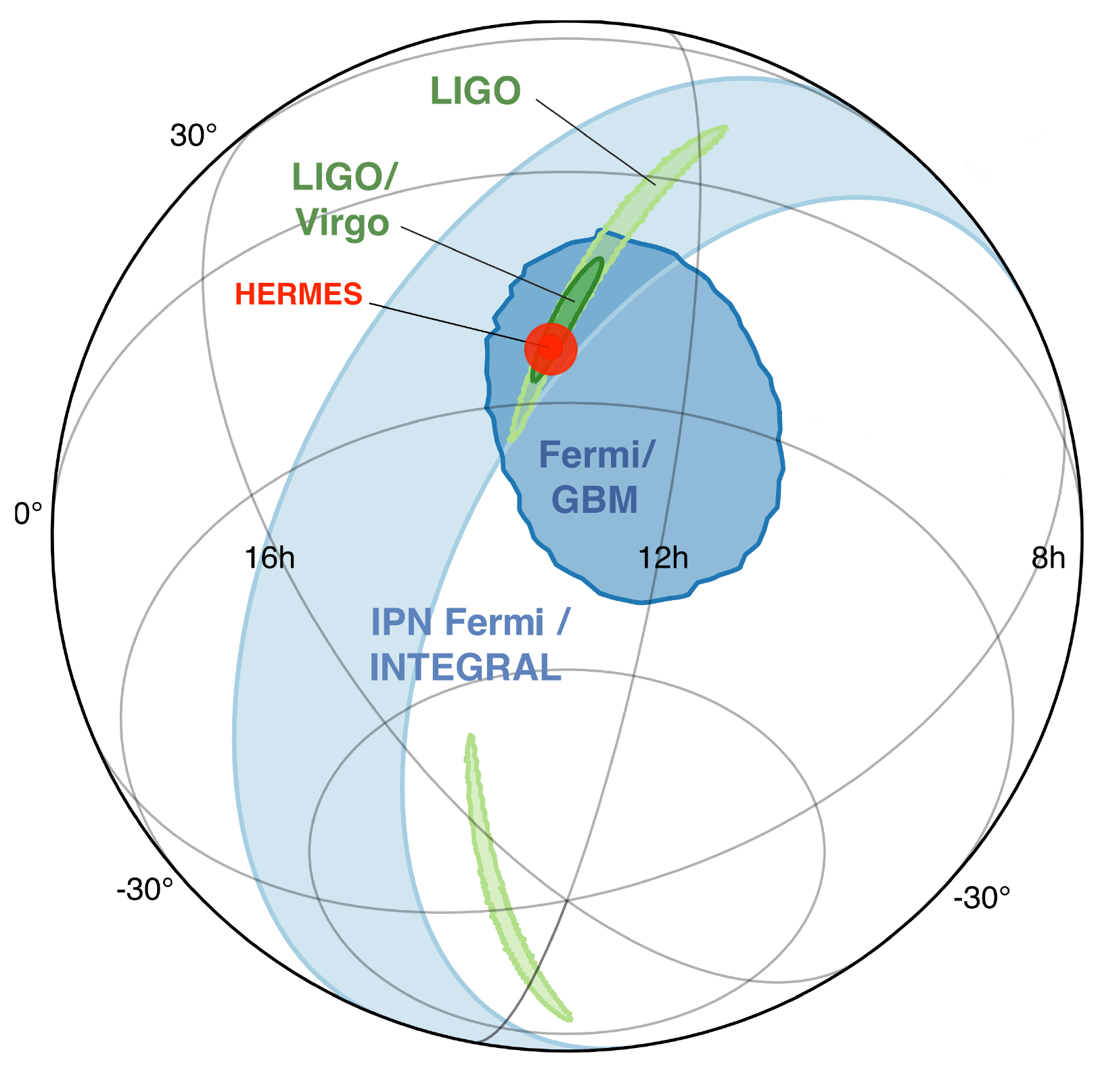} 
\end{tabular}
\end{center}
\caption[example] 
%>>>> use \label inside caption to get Fig. number with \ref{}
{\label{fig:gw170817} 
Accuracy in the location capabilities of LIGO, LIGO + Virgo, {\it HERMES} (red dot, $\sim 3\, {\rm deg}$), Fermi GBM, INTEGRAL. Original Figure from [\citenum{Abbot_mu_2017}].}. 
\end{figure}

\noindent The best fit formula is:
\begin{equation} 
\label{eq:taccA} 
\sigma_{\rm cross} \sim  \sigma_{\Delta t} = 3.3 \, \mu{\rm s} \times \left( \frac{A}{1\, {\rm m}^2} \right)^{-0.58}.
\end{equation}
In terms of the number of collected photons $N$ (adopting the same $0.8/6.5 \sim 40\%$ overall background) the formula is:
\begin{equation} 
\label{eq:taccN} 
\sigma_{\rm cross} \sim  \sigma_{\Delta t} = 3.3 \, \mu{\rm s} \times \left( \frac{N}{3.7 \times 10^6} \right)^{-0.58}.
\end{equation}
Inserting equation (\ref{eq:taccA}) into equation (\ref{eq:posacc}), it is possible to obtain the positional accuracy in the celestial coordinates of the bright Long GRB considered, once the average baseline, the effective area of each detector, and their number is known. 
As an example, we considered the location accuracies of {\it HERMES} Pathfinder, composed of three detectors with $56 \, {\rm cm}^2$ effective area in the energy band $50-300$ keV,  with an average baseline of 6000 km in case of a standard Long GRB, for which we adopted the rather conservative assumption of 
$\sigma_{\Delta t} = 1 \, {\rm ms}$ (for comparison the accuracy for the bright Long GRB described above, adopting equation (\ref{eq:taccA}) is $0.07\, {\rm ms}$).
In Figure~\ref{fig:gw170817}, we show the predicted location accuracies obtained with {\it HERMES} Pathfinder ($\sim 3\, {\rm deg}$) in comparison with the accuracies obtained for GW170817 by LIGO and Virgo, and GRB170817A by Fermi-GBM, and INTEGRAL [\citenum{Abbot_mu_2017}].

\subsection{The HERMES Project} 

\begin{figure}[h!]
\begin{center}
\begin{tabular}{cc}
\includegraphics[height=6cm]{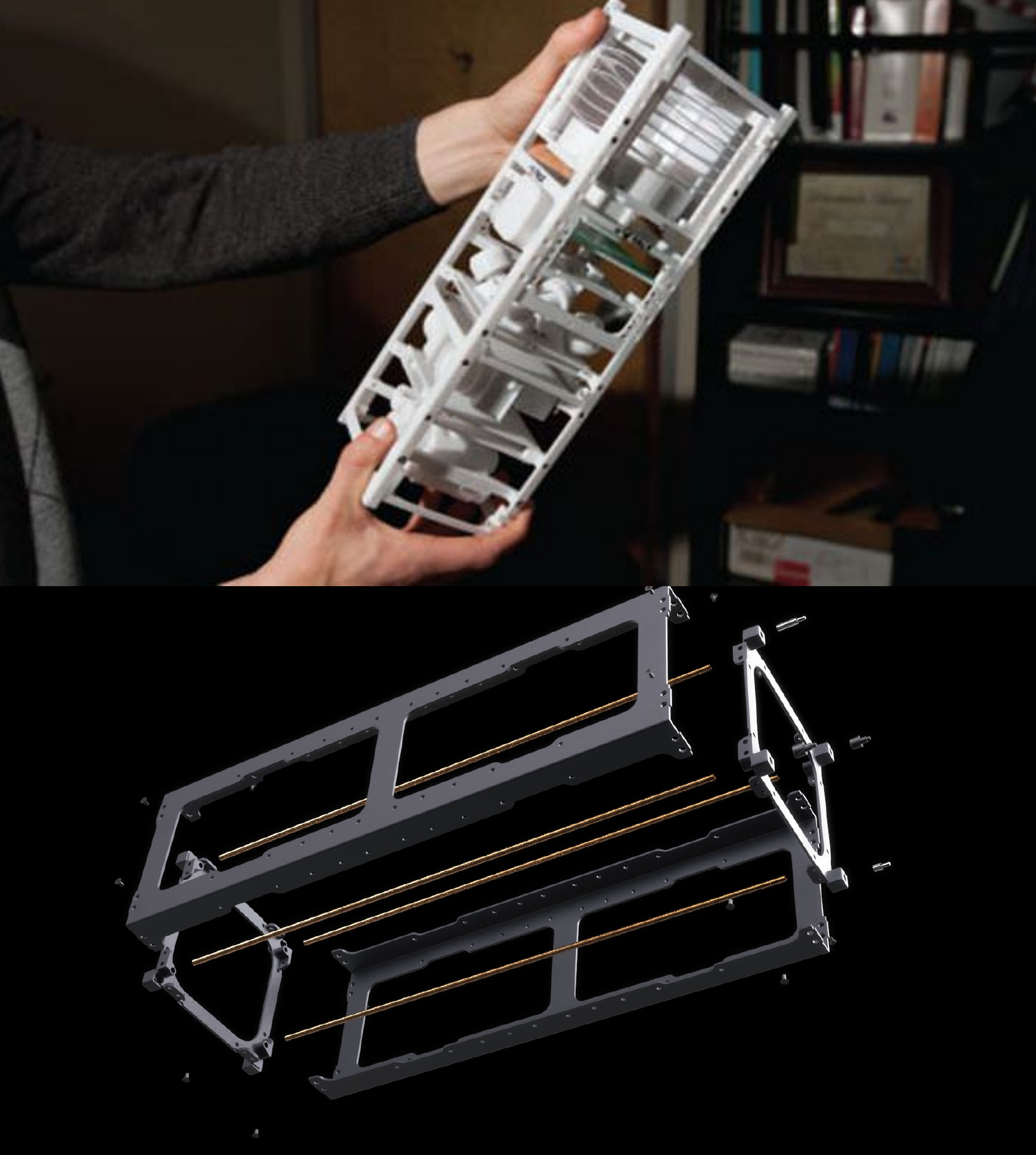} &
\includegraphics[height=6cm]{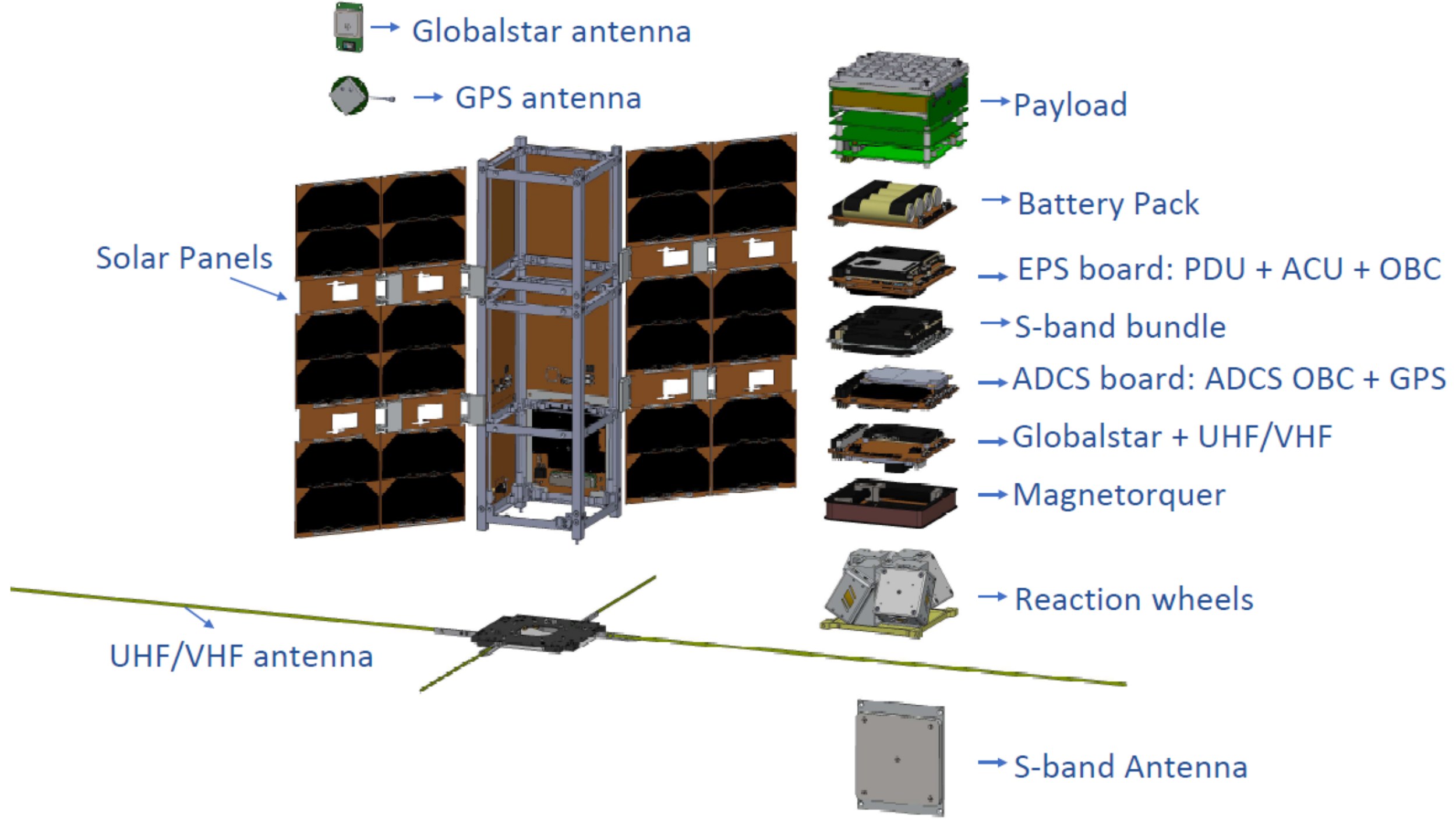} \\
Panel a) & Panel b) \\
\end{tabular}
\end{center}
\caption[example] 
%>>>> use \label inside caption to get Fig. number with \ref{}
{ \label{fig:spacecraft} 
Hermes 3U cubesat. 
\emph{Left panel)} Chassis of the spacecraft. Credit: Lawrence Livermore National Laboratory.
\emph{Right panel)} Exploded view of the spacecraft.}
\end{figure}

The {\it HERMES} Pathfinder is composed by a fleet of six 3U nano-satellites in equatorial Low Earth Orbit to be launched by the end of 2022. The structure of a 3U cube-sat is that of a parallelepiped $30 \times 10 \times 10 \, {\rm cm}$, which is the size of a champagne bottle. Figure~\ref{fig:spacecraft}, left panel, shows the chassis of the spacecraft, while the right panel, shows the exploded view of the spacecraft. The {\it HERMES} satellites have full gyroscopic stabilization and pointing capabilities. 
Data recording is continuous on internal buffer, and each satellite is equipped with S-band and VHF antennas for data download and command upload.

\begin{figure}[h!]
\begin{center}
\begin{tabular}{cc}
\includegraphics[height=6cm]{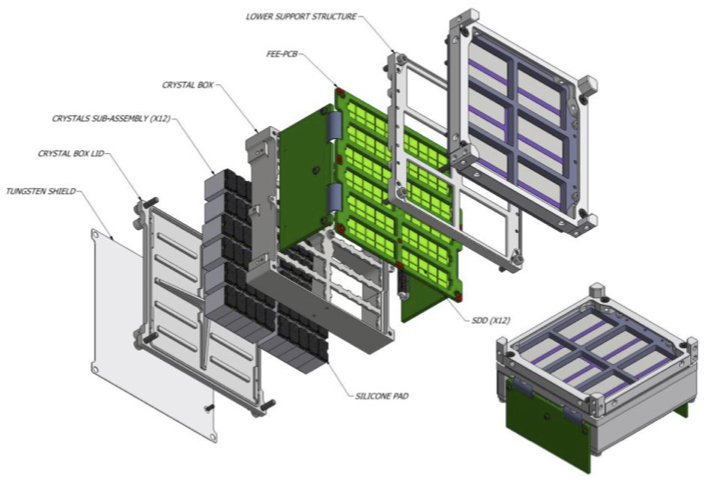} &
\includegraphics[height=6cm]{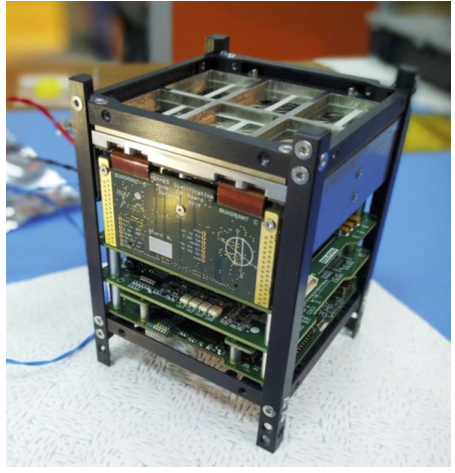} \\
Panel a) & Panel b) \\
\end{tabular}
\end{center}
\caption[example] 
%>>>> use \label inside caption to get Fig. number with \ref{}
{ \label{fig:payload} 
\emph{Left panel)} Exploded view of the payload. GAGGs are in grey and dark grey in the figure. SDD Array cells are the green squares.
\emph{Right panel)} Prototype of the HERMES detector.}
\end{figure}

The scintillator crystal X-/gamma-ray detector is located on top, with the detector window on the small face. 
It has a half-sky FoV (3 steradians FWHM). The solar panels, folded on the side of spacecraft, will be unfolded after the satellite release by means of the spring catapult.
Figure~\ref{fig:payload}, left panel,
shows the exploded view of the payload that is described in detail in an accompanying paper [\citenum{Evangelista20}].
Gadolinium-Aluminum-Gallium Garnet scintillator crystals (GAGGs, hereafter), in grey and dark grey in the figure, are arranged in blocks of five, for a total of twelve blocks (sixty crystals), for detection of photons in the band 20 keV $-$ 0.5 MeV. 
Each block is surmounted by a Silicon Drift Detector (SDD, hereafter) Array (each composed of 10 independent cells, green squares, in the figure) for direct reading of soft X-ray photons ($2-20$ keV) and scintillation optical photons from GAGGs. Passive shielding of the detector box is obtained by means of tungsten layers on bottom and sides to reduce X-ray and particle background. 
The total effective area is $56\, {\rm cm}^2$ and the temporal resolution is $\leq 0.5\, \mu{\rm s}$. 
%The mass is 1.8 kg. The volume is $10 \times 10 \times 12.5 \, {\rm cm}$. 
The current detector prototype is shown on Figure~\ref{fig:payload}, right panel. 

The {\it HERMES} Project has been fully financed by the Italian Space Agency (HERMES Technological Pathfinder) and the European community Horizon 2020 funds (HERMES Scientific Pathfinder) in the last four years, for a total amount just above eight million Euros. Launch and Operation costs (for a minimum of two years) will be supported by the Italian Space Agency.
 
\section{A shallow dive into Quantum Gravity: Minimal Length Hypothesis, Lorentz Invariance Violation, and Dispersion Relation for photons in vacuo} 

Several theories proposed to describe quantum Space-Time, for instance some String Theories, predict the existence of a minimal length for space of the order of Planck length, $\ell_{\rm PLANCK} = \sqrt{G \hbar/c^3} = 1.6 \times 10^{-33}\, {\rm cm}$ (see {\it e.g.} [\citenum{Hossenfelder12}] for a review).
This implies the following facts: 
\begin{itemize}
\item[]
{\it i)} these theories predict a Lorentz Invariance Violation (LIV, hereafter). According to Special Relativity, a proper length, $\ell$, is Lorentz contracted by a factor $\gamma^{-1} = [ 1 - (v/c)^2 ]^{1/2}$ when observed from a reference system moving at speed $v$ w.r.t. the reference system in which $\ell$ is at rest. If $\ell_{\rm MIN} = \alpha \, \ell_{\rm PLANCK}$ (where $\alpha \sim 1$ is a dimensionless constant that depends on the particular theory under consideration) is the minimal length physically conceivable (in String Theories $\ell_{\rm MIN}$ is the String length), no further Lorentz contraction must occur, at this scale. This is a violation of the Lorentz invariance.
\item[]
{\it ii)} These theories predict the remarkable fact that the space has, somehow, the structure of a crystal lattice, at Planck scale.
\item[]
{\it iii)} In perfect analogy with the propagation of light in crystals, these theories predict the existence of a dispersion law for photons {\it in vacuo} [\citenum{Amelino00}].
Since, for photons, energy scales as the inverse of the wavelength, this dispersion law can be expressed as a function of the energy of photons in units of 
the Quantum Gravity energy scale, which is the energy at which the quantum nature of gravity becomes relevant: $E_{\rm QG} = \zeta m_{\rm PLANCK} c^2 = \zeta
E_{\rm PLANCK}$, where $\zeta \sim \alpha^{-1} \sim 1$ is a dimensionless constant that depends on the particular theory under consideration, $m_{\rm PLANCK} = \sqrt{c \hbar/G} = 2.2 \times 10^{-5}\, {\rm g}$ is the Planck mass, and the Planck energy is $E_{\rm PLANCK} = 1.2 \times 10^{19}\, {\rm GeV}$:
\begin{equation} 
\label{eq:disp} 
|v_{\rm PHOT}/c - 1| \approx \alpha \left( \frac{E_{\rm PHOT}}{\zeta m_{\rm PLANCK} c^2} \right)^n
\end{equation}
where $\alpha \sim 1$ is a dimensionless constant that depends on the particular theory under consideration, $v_{\rm PHOT}$ is the group velocity of the photon wave-packet, and $E_{\rm PHOT}$ is the photon energy. The index $n$ is the order of the first relevant term in the expansion in the small parameter 
$\epsilon = E_{\rm PHOT}/(\zeta m_{\rm PLANCK} c^2)$. In several theories that predict the existence of a minimal length, typically,  $n=1$. Finally, the modulus is present in equation (\ref{eq:disp}) takes into account the possibility (predicted by different LIV theories) that higher energy photons are faster or slower than lower energy photons (discussed as sub-luminal, $+1$, or super-luminal, $-1$, as in [\citenum{Camelia09}].
\end{itemize}
We stress that not all the theories proposed to quantize gravity predict a LIV at some scale. This is certainly the case for Loop Quantum Gravity (see e.g. [\citenum{Rovelli88a,Rovelli90,Rovelli98}]). No LIV is expected as a consequence of the recently proposed Space-Time Uncertainty Principle [\citenum{Burderi16}] and in the Quantum Space-Time [\citenum{Sanchez2019}]. In some of these theories it is possible to conceive a photon dispersion relation that does not violate Lorentz invariance, although the first relevant term is quadratic in the ratio photon energy over $E_{\rm QG}$, {\it i.e.} $n=2$ in this case.
We explicitly note that, since $E_{\rm QG} \sim 10^{19}\, {\rm GeV}$, second order effects are almost not relevant even for photons of at $0.1 \, {\rm PeV}$ energies ($10^{14}\, {\rm eV}$), the highest energy photons ever recorded, recently confirmed to be emitted by the Crab Nebula [\citenum{Amenomori2019}]. Indeed also for these extreme photons $(E_{\rm PHOT}/E_{\rm QG})^2 \sim 10^{-28}$.

\subsection{Dispersion relation for photons in vacuo}
\label{sec:delays}

During motion at constant velocity, travel time is the ratio between the distance travelled $D_{\rm TRAV}$ and the speed. 
Therefore, differences in speed result in differences in the arrival times $\Delta t_{\rm QG}$ of photons of different energies $\Delta E_{\rm PHOT}$ departing from the same point at the same time, such as those emitted during a GRB. 
For small speed differences, as those predicted by the dispersion relations discussed above, these delays scales with the same order $n$ -- in the ratio 
$\Delta E_{\rm PHOT}/E_{\rm QG}$ -- as that between photon energy and Quantum Gravity energy scale: 
\begin{equation}
\label{eq:dtqg1}
\Delta t_{\rm QG} = \pm \xi  \, \left( \frac{D_{\rm TRAV}}{c} \right) \, \left( \frac{\Delta E_{\rm PHOT}}{\zeta m_{\rm PLANCK} \,c^2} \right)^n,
\end{equation}
where $\xi \sim 1$ is a dimensionless constant that depends on the particular theory under consideration and and the sign $\pm$ takes into account the possibility
(predicted by different LIV theories) that higher energy photons are faster or slower than lower energy photons respectively, as discussed above [\citenum{Camelia09}].

On the other hand, the distance traveled has to take into account the cosmological expansion, being a function of cosmological parameters and redshift. The comoving trajectory of a particle is obtained by writing its Hamiltonian in terms of the comoving momentum [\citenum{Jacob08}]. 
The computation of the delays has to take into account the fact that the proper distance varies as the universe expands. Photons of different energies are affected by different delays along the path, so, because of cosmological expansion, a delay produced further back in the path amounts to a larger delay on Earth.
Taking into account these effects this modified "distance traveled" $D_{\rm EXP}$ can be computed [\citenum{Jacob08}]. 

\noindent More specifically we adopted the so called Lambda Cold Dark Matter Cosmology ($\Lambda$CDM) with the following values [\citenum{Planck2016}]:
$H_{0} = 67.74(46) \, {\rm km\,s^{-1}Mpc^{-1}}$, 
$\Omega_{\rm k} = 0$, curvature $k=0$ that implies a flat Universe,  $\Omega_{\rm R} = 0$, radiation $= 0$ that implies a cold Universe,
$w = -1$, negative pressure Equation of State for the so called Dark Energy that implies an accelerating Universe,
$\Omega_{\Lambda} = 0.6911(62)$ and $\Omega_{\rm Matter} = 0.3089(62)$ (see [\citenum{Planck2016}], for the parameters and related uncertainties). \\
With these values we have:
\begin{equation}
\label{eq:dexp}
\frac{D_{\rm EXP}}{c} = \frac{1}{H_{0}} \int_0^z dz \frac{(1+z)}{\sqrt{\Omega_{\Lambda} +  
+ (1+z)^3\Omega_{\rm Matter}}},
\end{equation}
where $z$ is the redshift.

Substituting $D_{\rm TRAV}$ of equation (\ref{eq:dtqg1}) with $D_{\rm EXP}$ derived in equation (\ref{eq:dexp}) we finally obtain the delays between the time of arrival of photons of different energies as a function of the specific Dispersion Relation adopted, the specific Cosmology adopted, and the redshift:
\begin{equation}
\label{eq:dtqg2}
\Delta t_{\rm QG} = \pm \xi  \, \left( \frac{1}{H_{0}} \int_0^z dz \frac{(1+z)}{\sqrt{\Omega_{\Lambda} +  
+ (1+z)^3\Omega_{\rm Matter}}} \right) \, \left( \frac{\Delta E_{\rm PHOT}}{\zeta m_{\rm PLANCK} c^2} \right)^n.
\end{equation}

\subsection{Computation of the expected delays: long GRB at different redshifts}

We considered a bright Long GRB lasted for $\Delta t_{\rm GRB} = 40 \, {\rm s}$, with average flux in the 50-300 keV energy band 
$\phi_{\rm GRB} = 6.5 \, {\rm photons}/{\rm s}/{\rm cm}^2$,
background flux of $\phi_{\rm BCK} = 2.8 \, {\rm photons}/{\rm s}/{\rm cm}^2$, and variability timescale $\sim 5 \, {\rm ms}$ discussed in section \ref{sec:montecarlo}.
\begin{figure}[h!]
\begin{center}
\begin{tabular}{c}
\includegraphics[height=8cm]{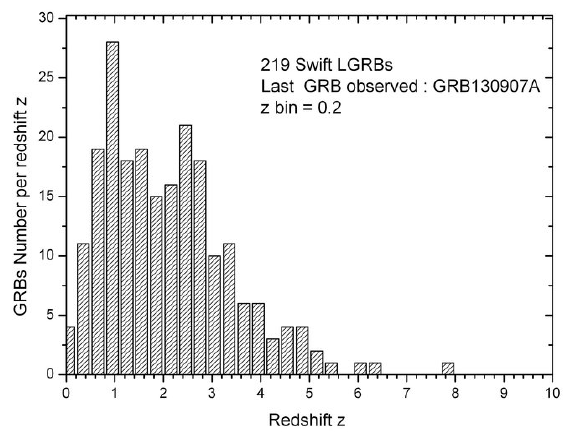} 
\end{tabular}
\end{center}
\caption[example]
%>>>> use \label inside caption to get Fig. number with \ref{}
{ \label{fig:redshifts} 
Distribution of 219 GRBs detected by Swift as a function of the redshift in bins of $\Delta z = 0.2$. Figure from [\citenum{Zitouni14}]
}
\end{figure}

We selected eight consecutive energy bands from 5 keV to 50 MeV. 
For an overall collecting area of $100\, {\rm m}^2$, the number of detected photons in each band was computed adopting a {\it Band} function, an empirical function that well fits GRB spectra [\citenum{Band93}]:
\begin{equation}
\label{eq:band}
\frac{dN_{E}(E)}{dA\; dt} =  F \times \left\{
\begin{array}{lr}
\left( \frac{E}{E_{\rm B}} \right)^{\alpha} \exp\{-(\alpha - \beta)E/E_{\rm B}\}, & E \le E_{\rm B}\\
\left( \frac{E}{E_{\rm B}} \right)^{\beta} \exp\{-(\alpha - \beta) \}, & E \ge E_{\rm B},
\end{array}
\right.
\end{equation}
where 
$E$ is the photon energy, 
$dN_{E}(E)/(dA\; dt)$ is the photon intensity 
energy distribution in units of ${\rm photons/cm^{2}/s/keV}$, 
$F$ is a normalization constant in units of ${\rm photons/cm^{2}/s/keV}$,
$E_{\rm B}$ is  the break energy, and $E_{\rm P} = [(2+\alpha)/(\alpha - \beta)] E_{\rm B}$ is the peak energy.
For most GRBs: $\alpha \sim -1$, $\beta \sim -2.0 \div -2.5$, and $E_{\rm B} \sim 225\; {\rm keV}$ that implies
$E_{\rm P} \sim 150\; {\rm keV}$.
We considered soft and hard cases ($\beta = -2.5$ and $\beta = -2.0$, respectively). Once the number of photons collected in each band $N$ is computed, the one sigma accuracy in the delays of the ToA of photons in a given energy band, 
$E_{CC}(N)$, is computed adopting the results of cross-correlation analysis performed on pairs of Monte-Carlo simulated GRBs in section \ref{sec:montecarlo} and expressed in equation (\ref{eq:taccN}) adopting the most conservative assumption that 
$E_{CC}(N)$ scales as $(N/3.7 \times 10^6)^{-0.5}$ (as expected from counting statistics) and not as $(N/3.7 \times 10^6)^{-0.58}$ of equation (\ref{eq:taccN}). 
We adopted the geometric mean of the lower and upper limits of a given energy band, $E_{\rm min}$ and $E_{\rm min}$ respectively, as representative of the average energy of the photons in that given band $E_{\rm AVE} = \sqrt{E_{\rm min} \times E_{\rm max}}$. With this, the energy difference between photons of different energy bands w.r.t. photons of very low energy $E_{\rm AVE} \sim 0$, are $\Delta E_{\rm PHOT} = E_{\rm AVE}$.
We adopted the cosmology described in section \ref{sec:delays}, a first order dispersion relation {\it i.e.} $n=1$, and, finally, $\xi =1$ and $\zeta = 1$.
The Quantum Gravity delays of the time of arrival of photons of different energy bands were computed with equation (\ref{eq:dtqg2}), for values of the redshift $z = 0.1, 0.5, 1.0, 3.0$, typical of GRBs as shown in Figure~\ref{fig:redshifts}. 
The results are shown in Table~\ref{table:1}. Numbers in red and blue refer to delays below and just above one sigma accuracy, respectively. 
Numbers in black are above three sigma.
\begin{table*}[h!]
%\vskip -1.2cm
\begin{center}
Quantum Gravity delays predicted with a first order photon dispersion relation
%Bright Long GRB: 8.00 (0.86 BCK) c/s ($50\div300$ keV) $-$  $\Delta t = 25$ s \\
%Spectral shape: {\it Band} function with $\alpha = -1$, $\beta = -2.5\div -2.0$, $E_{\rm B} = 225\; {\rm keV}$ \\
%Detector effective area: A$\,=100$ m$^2$ \\
%Accuracy in cross--correlation in function of the number of photons: $E_{CC}(N) = 0.46 \, \mu{\rm s} \sqrt{2.6\; 10^{8}/N}$ \\
%$\Lambda$CDM cosmology: $\Omega_{\Lambda} = 0.6911$ and $\Omega_{\rm Matter} = 0.3089$
\vskip 0.2cm
\bf{
\begin{tabular}{crccccrrrrr}
\hline
\hline
\\
\multicolumn{1}{c}{${\rm Energy \; band}$}    & \multicolumn{1}{c}{$E_{\rm AVE}$} & \multicolumn{1}{c}{$N$} &
\multicolumn{1}{c}{$E_{CC}(N)$} & 
\multicolumn{1}{c}{$N$}  & 
\multicolumn{1}{c}{$E_{CC}(N)$} & 
\multicolumn{4}{c}{${\rm \Delta t_{\rm QG}\; (\xi = 1.0, \; \zeta = 1.0)}$} \\ 
 & & $(\beta = -2.5)$ & & $(\beta = -2.0)$ & & & & & \\
\multicolumn{1}{c}{${\rm MeV}$}    & \multicolumn{1}{c}{${\rm MeV}$} & \multicolumn{1}{c}{${\rm photons}$} & 
\multicolumn{1}{c}{$\mu{\rm s}$} &
\multicolumn{1}{c}{${\rm photons}$}  & 
\multicolumn{1}{c}{$\mu{\rm s}$} &
\multicolumn{1}{c}{$\mu{\rm s}$} & \multicolumn{1}{c}{$\mu{\rm s}$} & 
\multicolumn{1}{c}{$\mu{\rm s}$} & \multicolumn{1}{c}{$\mu{\rm s}$} \\
 & & & & & &
\multicolumn{1}{c}{$z=0.1$} &         
\multicolumn{1}{c}{$z=0.5$} & \multicolumn{1}{c}{$z=1.0$} & \multicolumn{1}{c}{$z=3.0$} \\
\\
$0.005 - 0.025$ & $0.0112$ & $3.80 \times 10^8$ & $0.38$ & $3.02 \times 10^8$ & $0.43$ & {\color{red}$0.04$} & {\color{red}$0.25$} & {\color{red}$0.51$} & $1.42$ \\
$0.025 - 0.050$ & $0.0353$ & $1.40 \times 10^8$ & $0.62$ & $1.17 \times 10^8$ & $0.69$ & {\color{red}$0.13$} & {\color{red}$0.72$} & $1.46$ & $4.10$ \\
$0.050 - 0.100$ & $0.0707$ & $1.10 \times 10^8$ &  $0.71$ & $9.98 \times 10^7$ & $0.74$ & {\color{red}$0.27$} & $1.43$ & $2.93$ & $8.21$ \\
$0.100 - 0.300$ & $0.1732$ & $8.98 \times 10^7$ & $0.79$ & $1.00 \times 10^8$ & $0.74$ & {\color{red}$0.66$} & $3.51$ & $7.19$ & $20.10$ \\
$0.300 - 1.000$ & $0.5477$ & $2.07 \times 10^7$ & $1.64$ & $3.82 \times 10^7$ & $1.20$ & {\color{red}$2.09$} & $11.11$ & $22.72$ & $63.56$ \\
$1.000 - 2.000$ & $1.4142$ & $2.63 \times 10^6$ & $4.56$ & $8.20 \times 10^6$ & $2.60$ & {\color{blue}$5.40$} & $28.68$ & $58.67$ & $164.12$ \\
$2.000 - 5.000$ & $3.1623$ & $1.07 \times 10^6$ & $7.19$ & $4.92 \times 10^6$ & $3.35$ & {\color{blue}$12.07$} & $64.12$ & $131.19$ & $367.00$ \\
$5.000 - 50.00$ & $15.8114$ & $3.52 \times 10^5$ & $12.54$ & $2.95 \times 10^6$ & $4.33$ & $60.35$ & $320.62$ & $656.00$ & $1834.98$ \\
\\
\hline
\hline
\end{tabular}
}
\end{center}
\caption{\label{table:1} Photon fluence and expected delays induced by a Quantum Gravity first order Dispersion Relation for the bright 
Long GRB described in section \ref{sec:montecarlo} and observed with a detector of cumulative 
effective area of $100\; {\rm m}^2$ ({e.g.} obtained by adding the photons collected by $N = 10^{4}$ nano--satellites of $100\; {\rm cm}^2$ each). 
The GRB is described by a {\it Band function} 
with $\alpha = -1$, $\beta = -2.5 \div -2.0$, $E_{\rm B} \sim 225\; {\rm keV}$. 
The modified "distance traveled" by the photons $D_{\rm EXP}$ described in the text has been computed for each redshift
adopting a $\Lambda$CDM cosmology with $\Omega_{\Lambda} = 0.6911$ and $\Omega_{\rm Matter} = 0.3089$. 
This implies the following:
$D_{\rm EXP} = 453.9\, {\rm Mpc}$ for $z=0.1$, 
$D_{\rm EXP} = 2411.4\, {\rm Mpc}$ for $z=0.5$,
$D_{\rm EXP} = 4933.6\, {\rm Mpc}$ for $z=1.0$, $D_{\rm EXP} = 13801.2\, {\rm Mpc}$ for $z=3.0$.
Adopting $n=1$, $\xi =1$ and $\zeta =1$, we found 
$| \Delta t_{\rm QG} | = 3.8168\, {\rm \mu s} \times \Delta E_{\rm PHOT}/ (1\; {\rm MeV})$ for $z=0.1$, 
$| \Delta t_{\rm QG} | = 20.2775\, {\rm \mu s} \times \Delta E_{\rm PHOT}/ (1\; {\rm MeV})$ for $z=0.5$, 
$| \Delta t_{\rm QG} | = 41.4863\, {\rm \mu s} \times \Delta E_{\rm PHOT}/ (1\; {\rm MeV})$ for $z=1.0$, $| \Delta t_{\rm QG} | = 116.0544\, {\rm \mu s} \times \Delta E_{\rm PHOT}/ (1\; {\rm MeV})$ for $z=3.0$. $\Delta E_{\rm PHOT} = E_{\rm AVE} = \sqrt{\rm E_{max} \times E_{min}}$ (see text). The (statistical) cross--correlation accuracies are computed as 
$E_{CC}(N) = 3.3 \, \mu{\rm s} \sqrt{3.7\; 10^{6}/N}$, obtained from Monte--Carlo simulations.
}
\end{table*}

\subsection{Intrinsic delays or Quantum Gravity delays?}

Because of unknown details on the {\it Fireball model}, intrinsic delays in the emission of photons of different energy bands are possible. For a given GRB, these intrinsic delays can mix to, or even mimic, a genuine quantum gravity effect, making its detection impossible. However, intrinsic delays in the emission mechanism are independent of the distance of the GRB. 
On the other hand, the delays induced by a photon dispersion law are proportional both to the distance traveled (known function of redshift) and to the differences in energy of the photons. 
This double dependence on energy and redshift is the unique signature of a genuine Quantum Gravity effect.
This behavior, shown in Table~\ref{table:1}, demonstrates that, given an adequate collecting area, GRBs are indeed excellent tools to effectively search for a first order dispersion law for photons, once their redshifts are known. 
 
%\section{Location of GRBs  with fleets of satellites and redshifts measurements}
\section{GRB localization and redshift measurements}

Distributed astronomy offers a double vantage for detecting transient events in the high-energy sky: 
\begin{itemize}
\item[]
{\it i)} thanks to the possibility of reaching an overall huge collection area, it allows to reach extraordinary sensitivity and to collect an impressive number of photons, resulting in high statistics even at tiny temporal scales;
\item[]
{\it ii)} by means of temporal triangulation techniques, it allows for unprecedented accuracies in location capabilities for an all-sky monitor with half-sky field of view and no pointing capabilities. 
\end{itemize}
The accuracy in locating the prompt emission of GRBs is particularly relevant as it allows for fast follow-up from large optical telescopes and 
determination, in almost all cases, of the redshift of the host galaxy. As an example, we consider the bright Long GRB described in section \ref{sec:montecarlo}, for which we compute the positional accuracy for the following configuration of satellites:
\begin{itemize}
\item[]
{\it a)} Large fleet of small satellites in Low Earth Orbits: \\
$A = 30 \times 30\, {\rm cm} \approx 0.1 {\rm m}^2$ that implies
$\sigma_{\Delta t} = 12.5\, \mu{\rm s}$ \\
Average baseline $D \approx 6,000 \,{\rm km}$ \\
$N_{\rm DET} \approx 1000$ \\
$\sigma_{\alpha \, {\rm STAT}} \sim  \sigma_{\delta \, {\rm STAT}} \approx 4\, {\rm arcsec}$ 
\item[]
{\it b)} Three satellites with large detectors in Earth-Moon system Lagrangian points: \\
$A = 1 {\rm m}^2$ that implies $\sigma_{\Delta t} = 3.3\, \mu{\rm s}$. \\
Average baseline $D \approx 400,000 \,{\rm km}$ \\
$N_{\rm DET} = 3$ \\
$\sigma_{\alpha \, {\rm STAT}} \sim  \sigma_{\delta \, {\rm STAT}} \approx 0.5\, {\rm arcsec}$
\end{itemize} 

\begin{comment}
\section{GrailQuest: First Quantum-Gravity dedicated experiment}

We conceived {\it GrailQuest} as the first large astrophysical experiment dedicated to Quantum Gravity. 
The main objective of this experiment is the effective search for a first order dispersion law for photons in vacuo to explore Space-Time structure down to the Planck scale. 

We demonstrated that this ambitious goal is possible with an all-sky monitor of the Gamma-ray sky (50 keV -- 50 MeV energy band) distributed in space, with overall collecting area of the order of several tens of square meters, and very fast time resolution of $\sim 0.1 \, \mu{\rm s}$. 
Crystal scintillators read by Silicon Photomultiplier or Silicon Drift Detectors are a promising class of detectors for this experiment, under study at present moment. 
Temporal triangulation techniques allow to locate GRBs within few arc-seconds, allowing fast follow-up with optical telescope to obtain redshifts. 

In order to promote the potential of Distributed Astronomy and to support the {\it GrailQuest} project, we submitted a white paper in response to an European Space Agency call for the scientific long term plan Voyage 2050, following the last plan, Cosmic Vision, started in 2004. 
Voyage 2050, will cover the period from 2035 to 2050. 
The paper has been accepted to be published in a dedicated issue of {\it Experimental Astronomy} [\citenum{Burderi2019}]. \\

\end{comment}

\section{GrailQuest: First Quantum-Gravity dedicated experiment}
%\section{GrailQuest Next Generation: a symbiont on satellites of Mega-Constellations}

We conceived {\it GrailQuest} as the first large astrophysical experiment dedicated to Quantum Gravity. 
The main objective of this experiment is the effective search for a first order dispersion law for photons in vacuo to explore Space-Time structure down to the Planck scale. 

We demonstrated that this ambitious goal is possible with an all-sky monitor of the Gamma-ray sky (50 keV -- 50 MeV energy band) distributed in space, with overall collecting area of the order of several tens of square meters, and very fast time resolution of $\sim 0.1 \, \mu{\rm s}$. 
Crystal scintillators read by Silicon Photomultiplier or Silicon Drift Detectors are a promising class of detectors for this experiment, under study at present moment. 
Temporal triangulation techniques allow to locate GRBs within few arc-seconds, allowing fast follow-up with optical telescope to obtain redshifts. 

In order to promote the potential of Distributed Astronomy and to support the {\it GrailQuest} project, we submitted a white paper in response to an European Space Agency call for the scientific long term plan Voyage 2050, following the last plan, Cosmic Vision, started in 2004. 
Voyage 2050, will cover the period from 2035 to 2050. 
The paper has been accepted to be published in a dedicated issue of {\it Experimental Astronomy} [\citenum{Burderi2019}]. \\
%Please, download paper from arXiv (arXiv:1911.02154v2). 

A compelling possibility for the future of distributed astronomy and, in particular, for the {\it GrailQuest} project, is to host the detectors, as symbionts, on the large constellations of satellites on Earth orbit. These constellations (mega-constellations, hereafter) are already under construction, or planned for the immediate future, to provide satellite internet access worldwide. 

{\it OneWeb}\footnote{\url{https://www.oneweb.world/}} is a constellation of 650 satellites (150 kg each), in a circular Earth orbit, at 1,200 km altitude, owned, among others, by Virgin Galactic, Arianespace, and Airbus Defence and Space. They started launching satellites in 2019 and at present already 110 satellites are operational. They recently planned to increase the number of satellites up to several thousands. \\
Boarding a $30 \times 30 \, {\rm cm}$ effective area detectors on each of the originally planned satellites, would result in a $\sim 60 \, {\rm m}^2$ effective area all-sky monitor.

{\it Starlink}\footnote{\url{https://www.starlink.com/}} is a constellation of satellites (12,000 approved by International Telecommunication Union plus 30,000 requested and under approval) under construction by Space-X. The satellites (between 100 and 500 kg each) will be deployed in circular orbits between 340 and 1,100 km altitude. Launches started in 2018 and about 1,000 satellites were launched up to date. Even boarding a small $10 \times 10 \, {\rm cm}$ effective area detector on each of the originally planned satellites, would result in a $\sim 120 \, {\rm m}^2$ effective area all-sky monitor.

{\it Kuiper System}\footnote{\url{https://www.amazon.jobs/en/teams/projectkuiper}}, an Amazon project, is a planned constellation of 3,000 satellites in circular orbits at 600 km altitude, proposed in 2019. Also in this case, the detectors of the {\it GrailQuest} project could be hosted as symbionts on the satellites of this constellation. 

In the name of scientific progress, the companies that are constructing the mega-constellations could bear part of the costs of building the detectors and managing the flow of scientific data. 
Indeed, philanthropy has often aided the advancement of science and astronomy in particular. The famous Hale reflector in Palomar Observatory (5.1 m in diameter) was funded by Rockefeller Foundation in 1928 on a proposal by the astronomer George Ellery Hale. 

The last point we want to highlight, here, is another fundamental aspect of distributed astronomy which is that of being modular and scalable through the replication of identical and easy to implement detectors. This could allow a sort of mass production with a massive cut in production costs, in line with the great intuition of Henry Ford, inventor of the assembly line. 
A quantum leap for astronomy of the third millennium.

\section{GrailQuest: conclusions}

\begin{figure}[h!]
\begin{center}
\begin{tabular}{c}
\includegraphics[height=12cm]{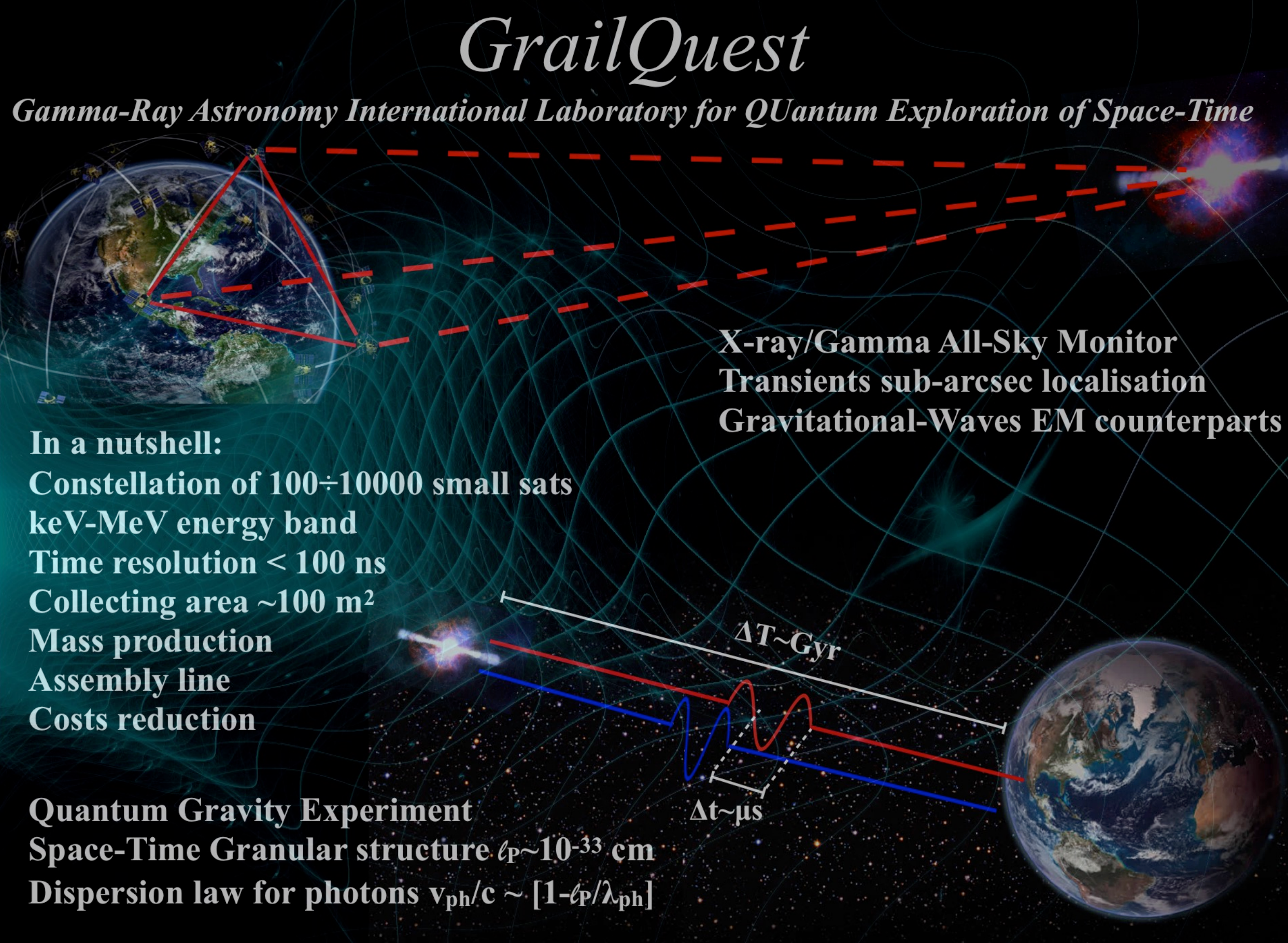} 
\end{tabular}
\end{center}
\caption[example]
%>>>> use \label inside caption to get Fig. number with \ref{}
{ \label{fig:grailquest} 
The {\it GrailQuest} project.
}
\end{figure}

Main conclusions on the {\it GrailQuest} project are shown in Figure~\ref{fig:grailquest} and summarized in the following points:
\begin{itemize}
\item[]
{\it i)}
{\it GrailQuest} is a modular astrophysical observatory hosted on hundreds/thousands small satellites. 
\item[]
{\it ii)} The simultaneous use of very small and simple sub-units allow to reach a huge overall collecting area (hundreds of square meters). 
\item[]
{\it iii)} {\it GrailQuest} is an all-sky monitor for transient events in the X-/gamma-ray band (from few keV to 50 MeV). 
\item[]
{\it iv)} Extraordinary temporal resolution ($0.1\, \mu{\rm s}$) allows, with temporal triangulation techniques, to localize the events down to sub-arcsecond accuracies.
\item[]
{\it v)} {\it GrailQuest} will be the perfect hunter for the electromagnetic counterparts of Gravitational Wave Events. 
\item[]
{\it vi)} {\it GrailQuest} will perform the first large scale dedicated Quantum Gravity experiment to search for a first order (in the ratio photon energy over Quantum Gravity energy) dispersion law for photons {\it in vacuo}, constraining the Space granular structure down to the minuscule Planck length, 
$\ell_{\rm PLANCK} = \sqrt{G \hbar/c^3} = 1.6 \times 10^{-33}\, {\rm cm}$.
\item[]
{\it vii)} Mass production of each module (Assembly Line philosophy) will allow huge reduction of costs.
\end{itemize}

\acknowledgments % equivalent to \section*{ACKNOWLEDGMENTS}       
 
This work has been carried out in the framework of the HERMES-TP and HERMES-SP collaborations. We acknowledge support from the European Union Horizon 2020 Research and Innovation Framework Program under grant agreement HERMES-Scientific Pathfinder n. 821896 and from ASI-INAF Accordo Attuativo HERMES Technologic Pathfinder n. 2018-10-H.1-2020. LB and AS acknowledge financial contribution from the PRIN 2017 agreement n. 20179ZF5KS.  LB, AS and AR acknowledge financial contribution from the FdS 2017, CUP n. F71I17000150002. \\

% References
%\bibliography{report} % bibliography data in report.bib
\bibliography{biblio}
\bibliographystyle{spiebib} % makes bibtex use spiebib.bst

\end{document}